\documentclass[12pt]{article}
\usepackage{graphicx,amsmath,amssymb}
\usepackage{units}
\usepackage{color}

\newlength{\dinwidth}
\newlength{\dinmargin}
\def\lapproxeq{\lower .7ex\hbox{$\;\stackrel{\textstyle                                                    
<}{\sim}\;$}}                                                    
\def\gapproxeq{\lower .7ex\hbox{$\;\stackrel{\textstyle                                                    
>}{\sim}\;$}}                                                    
\def\be{\begin{equation}}                                                    
\def\ee{\end{equation}}                                                    
\def\bea{\begin{eqnarray}}                      
\def\eea{\end{eqnarray}}

\def\GeV{\rm GeV}
\def\TeV{\rm TeV}

\def\sh{\hat s}
\def\sh2{{\hat s}^2}

\begin{document}

\begin{flushright}                                                    
LCTS/2015-45  \\
IPPP/15/81  \\
DCPT/15/162 \\                                                    
\today \\                                                    
\end{flushright} 

\vspace*{0.5cm}

\begin{center}
{\Large \bf The impact of the final HERA combined data}\\ 
\vspace*{0.5cm}{\Large \bf on PDFs obtained from a global fit}\\

\vspace*{1cm}
L. A. Harland-Lang$^{a}$, A. D. Martin$^b$, P. Motylinski$^a$ and R.S. Thorne$^a$\\                                               
\vspace*{0.5cm}                                                    
                                                  
$^a$ Department of Physics and Astronomy, University College London, WC1E 6BT, UK \\           
$^b$ Institute for Particle Physics Phenomenology, Durham University, DH1 3LE, UK                                                    
                                                    
\vspace*{1cm}

\begin{abstract}
\noindent We investigate the effect of including the HERA run I + II 
combined cross section data on the MMHT2014 PDFs. We present the fit quality
within the context of the global fit and when only the HERA data are included.  
We examine the changes in both the central values and uncertainties in the 
PDFs. We find that the prediction for the data is good, and only relatively 
small improvements in $\chi^2$ and changes in the PDFs are obtained 
with a refit at both NLO and NNLO. PDF uncertainties are slightly reduced. There is a small
dependence of the fit quality on the value of $Q^2_{\min}$. This can be 
improved by phenomenologically motived corrections to $F_L(x,Q^2)$ 
which parametrically are largely in the form of higher--twist type 
contributions. 

\end{abstract}
                                                   
\vspace*{0.5cm}                                                    
                                                    
\end{center}

\section{Introduction}\label{sec:intro} 

The MSTW2008 PDFs \cite{MSTW} have been widely used in the analyses 
of hadron collider data. They were recently updated with an 
analysis performed in the same general framework, resulting in the 
the MMHT2014 PDFs \cite{Harland-Lang:2014zoa}, and accompany recent updates 
by other groups \cite{ABM14,JR14,NNPDF3,CT14}, with the CT, MMHT and 
NNPDF sets having been combined in an updated PDF4LHC recommendation
\cite{Butterworth:2015oua}.
The MMHT 2014 PDFs were an improvement to the MSTW 2008 PDFs partially
due to a number of developments in the
procedures employed in the analysis. 
For example, we now use modified and extended parameterisations for the 
PDFs based on Chebyshev
polynomials, and we allow freedom in the deuteron nuclear corrections, 
both these features being 
introduced in \cite{MMSTWW}. This led to a change in
the $u_V-d_V$ distribution and an improved description of the LHC data
for the $W$ boson charge asymmetry.
Additionally, we now use the ``optimal'' GM-VFNS choice \cite{Thorne}
which is smoother near
to heavy flavour transition points, particularly at NLO.
The correlated systematic uncertainties, which are important for jet data
in particular, are now treated as multiplicative rather than
additive. We have also changed the
value of the charm branching ratio to muons used to $B_{\mu} = 0.092$ and 
allow an uncertainty of $\pm 10\%$
\cite{Bolton}. This feeds into the central value and the uncertainty 
of the strange quark PDF.

There are also a wide variety of
new data sets included in the MMHT fit.
These include $W,Z$ cross sections from ATLAS, CMS and LHCb, 
differential in rapidity; Drell Yan data at high and low mass; and
also data on $\sigma_{t\bar t}$ from the Tevatron and from ATLAS and CMS.
At NLO we also include ATLAS and CMS inclusive jet data from the 7~TeV run, 
though we do not yet include these data at NNLO. Previous
analyses have used threshold corrections for the Tevatron jet data, and we continue
to include these data in the NNLO analysis. However for jet data from the
LHC we are often far from threshold, and the approximation to the full 
NNLO calculation is not likely to be reliable.
The full
NNLO calculation \cite{GGGP1,GGGP2} is nearing
completion.
There are also various changes in non-LHC data sets, for example we include some 
updated Tevatron $W$ boson asymmetry data sets.
The single most important change in data included is the
replacement of the HERA run I neutral and charged current data provided
separately by H1 and ZEUS with the combined HERA data set \cite{H1+ZEUS}
(and we also include HERA combined data on $F_2^c(x,Q^2)$
\cite{H1+ZEUScharm}). These are the data which provide
the best single constraint on PDFs, particularly on the gluon at all
$x < 0.1$.

However, in \cite{Harland-Lang:2014zoa} we decided not to include any 
separate run II H1 and ZEUS data sets
since it was clear the full run I $+$ II combined data would soon
appear. This has now recently happened, and the data, and the accompanying
PDF analysis, are  published in \cite{Abramowicz:2015mha}.
It was not stated in \cite{Harland-Lang:2014zoa} precisely when
an update of MMHT2014 PDFs would be required. Significant new LHC data would
be one potential reason, and the full NNLO calculation of the jet
cross sections, effectively allowing a larger data set at NNLO,  
might be another. The potential impact of the final
HERA inclusive cross section data was another factor in this 
decision, it being possible that these
alone might produce a very significant change in either 
the central value of the PDFs
or their uncertainties, or both. Hence, it is now obviously a high priority to
investigate their impact.\footnote{Initial results were presented in 
\cite{Thorne:2015caa} and similar results were also found in 
\cite{Rojo:2015nxa}.} However, as well as just investigating the impact of
the new data on the PDFs assuming a standard fixed-order perturbative 
treatment, it is also interesting to investigate the quality of the fit, and 
to see if it is possible to improve the quality in some regions of $x$ and 
$Q^2$. In particular, there is a suggestion in \cite{Abramowicz:2015mha}
that the data at low $Q^2$ are not fit as well as they could be, so we 
first confirm that we also see this feature, and also investigate, in a very 
simple manner, what type of corrections can solve this problem.

\section{Fit to combined HERA data set}\label{sec:herafit}

If we use our standard cut of $Q^2_{\min}=2~\GeV^2$ to eliminate data with $Q^2$
below this value, there are 1185 HERA 
data points 
with 162 correlated systematics and 7 procedural uncertainties. 
These are naturally separated into 7 subsets, depending on whether 
the data are obtained from $e^+$
or $e^-$ scattering from the proton, whether it is from neutral or charged 
current scattering, and on the proton beam energy $E_p$. 
This is to be compared to 621 data points, separated into 5 subsets, 
with generally larger uncertainties, from the HERA I combined data used 
previously (though these data do have fewer correlated systematics).  
We first investigate the fit quality from the predictions 
using MMHT2014 PDFs and without performing any refit. We use the same 
$\chi^2$ definition as in \cite{Harland-Lang:2014zoa}, i.e. 
\be
\chi^2=\sum_{i=1}^{N_{\rm pts}}\left(\frac{D_i+\sum_{k=1}^{N_{\rm corr}}
r_k\sigma_{k,i}^{\rm corr}-T_i}{\sigma_i^{\rm uncorr}}\right)^2+\sum_{k=1}^{N_{\rm corr}}r_k^2,
\ee 
where $D_i+\sum_{k=1}^{N_{\rm corr}}r_k\sigma_{k,i}^{\rm corr}$ are the data 
values allowed to shift by some multiple $r_k$ of the systematic error
$\sigma_{k,i}^{\rm corr}$ in order to give the best 
fit, and where $T_i$ are the parametrised predictions. 
The results obtained are already rather good:

\medskip

\noindent\centerline{$\chi^2_{\rm NLO} = 1611/1185 = 1.36$ per point.}

\noindent\centerline{$\chi^2_{\rm NNLO} = 1503/1185 = 1.27$ per point.}

\medskip

\noindent This is to be compared to the result in \cite{Abramowicz:2015mha} 
with HERAPDF2.0 PDFs, which are fit to (only) these data. They  
obtain $\sim 1.20$ per point using $Q^2_{\min}=2~\GeV^2$, at both NLO and 
NNLO. Hence, we do not expect dramatic improvement to the fit quality
from our predictions by refitting, particularly at NNLO. 
Next we perform a refit in the context of our standard global fit, i.e. we 
simply replace the previous HERA run I data with the 
new run I $+$ II combined data. There are no procedural changes to the fit 
at all. The fit quality improves to   

\medskip

\noindent\centerline{$\chi^2_{\rm NLO} = 1533/1185 = 1.29$ per point, with
deterioration $\Delta \chi^2 = 29$ in other data.} 

\noindent\centerline{$\chi^2_{\rm NNLO} = 1457/1185 = 1.23$ per point, with
deterioration $\Delta \chi^2 = 12$ in other data.} 

\medskip

\noindent This is a significant, but hardly dramatic improvement (and much 
less than
the improvement after refitting when HERA run I combined data were first 
introduced into the MSTW2008 fitting framework \cite{MSTWDIS}), i.e. 
the MMHT2014 PDFs are
already giving quite close to the best fit within the global fit framework.  

\begin{table}[h]
\begin{center}
\begin{tabular}{|l|c|c|c|c|c|}
\hline
          & no. points    & NLO $\chi^2_{\rm HERA}$  &   NLO $\chi^2_{\rm global}$ &
  NNLO $\chi^2_{\rm HERA}$  &   NNLO $\chi^2_{\rm global}$         \\
\hline
correlated penalty      &       & 79.9 &  113.6& 73.0 &  92.1  \\
CC $e^+p$                &  39   & 43.4 & 47.6  & 42.2 & 48.4     \\
CC $e^-p$                &  42   & 52.6 & 70.3  & 47.0 & 59.3     \\
NC $e^-p$ $E_p=920~\GeV$ &  159  & 213.6 & 233.1 & 213.5 & 226.7     \\
NC $e^+p$ $E_p=920~\GeV$ &  377  & 435.2 & 470.0 & 422.8 & 450.1   \\
NC $e^+p$ $E_p=820~\GeV$ &   70  & 67.6  & 69.8 & 71.2  & 69.5   \\
NC $e^-p$ $E_p=575~\GeV$ &   254 & 228.7 & 233.6 & 229.1 & 231.8     \\
NC $e^-p$ $E_p=460~\GeV$ &   204 & 221.6 & 228.1 & 220.2 & 225.6     \\
\hline                                                        
total      &      1145          &1342.6 &1466.1&1319.0 &1403.5  \\
\hline
    \end{tabular}
\vspace{-0.0cm}
\caption{\label{tab:t1} The $\chi^2$ for each subset of HERA I + II 
data for our four different fits with $Q^2_{\min}=3.5~\GeV^2$.  Note that this data cut
eliminates 40 HERA data points as compared to fit with $Q^2_{\min}=2~\GeV^2$. 
In this table the $\chi^2$ per data set does not include the penalties for 
shifts in systematic parameters, which is separated out at the top of the 
table. This is the only place in the article where this separation has been 
made.} 
\vspace{-0.4cm}
\end{center}
\end{table}

In order to compare more directly with the HERAPDF2.0 study we
also fit to only HERA run I $+$ II data. This requires us to 
fix 4 of our normally free PDF parameters in order to avoid 
particularly unusual PDFs.  In
practice the potential danger is a very complicated, and potentially 
pathological, strange quark distribution, which can fluctuate 
dramatically as
HERA data do not have any direct constraint on the $s$ and ${\bar s}$ PDFs. We allow the $s + \bar s$
distribution to have a free normalisation and high-$x$ power
but all other shape freedom is removed. The $s-\bar s$ asymmetry is 
fixed to the MMHT2014 default value. With these restrictions, the result of our fit is  

\medskip

\noindent\centerline{$\chi^2_{\rm NLO} = 1416/1185 = 1.19$ per point}

\noindent\centerline{$\chi^2_{\rm NNLO} = 1381/1185 = 1.17$ per point}
  
\medskip

\noindent Hence, in this case, as well as the global fit, the NNLO fit 
quality is still definitely better than that at NLO, but not as distinctly.

\begin{figure}
\begin{center}
\vspace*{-0.3cm}
\includegraphics[scale=0.4]{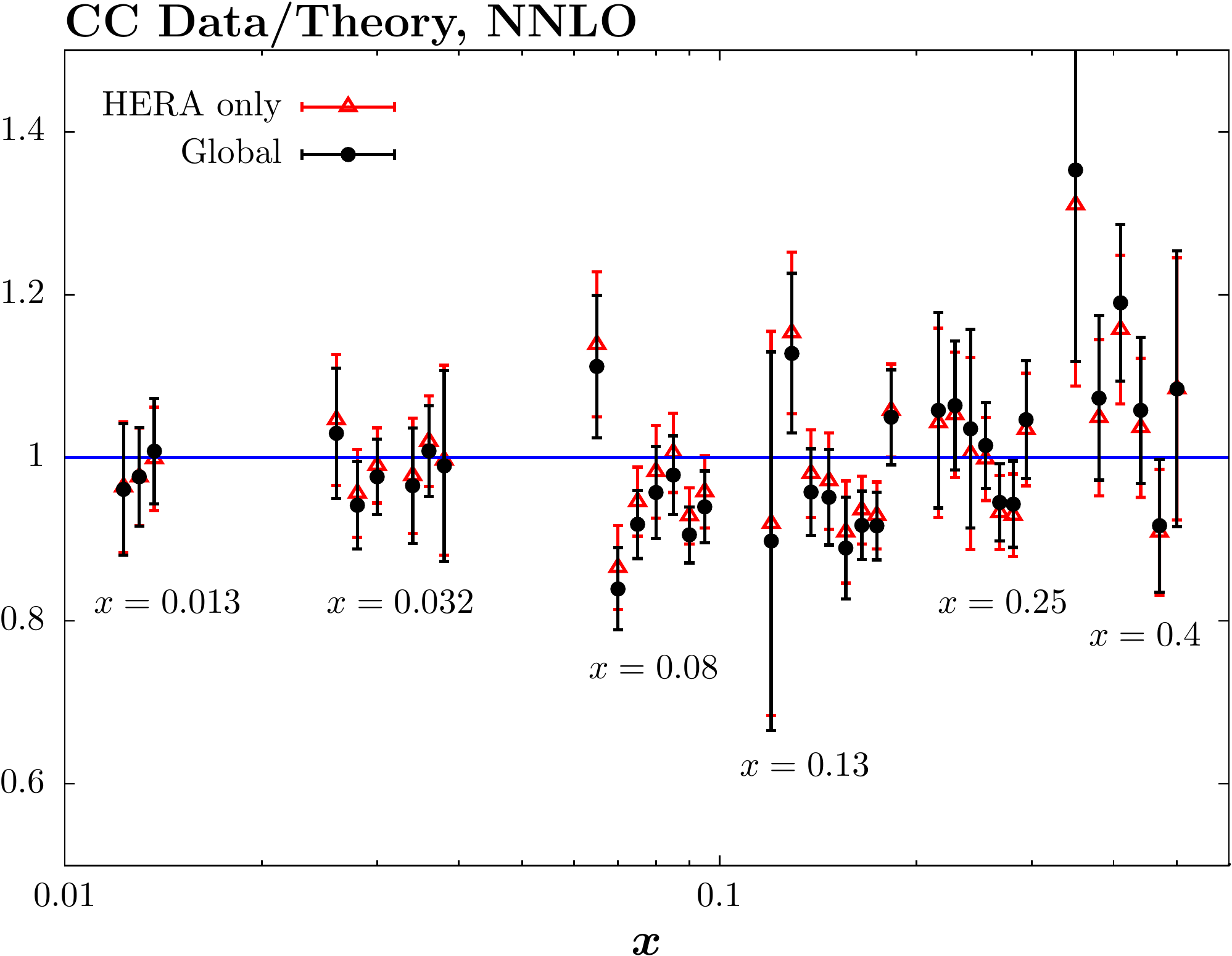}
\caption{\sf HERA $e^-$ charged current data divided by theory for the local fit to HERA II combined data, and for the global fit including this data set. 
The shifts of data relative to theory due to correlated uncertainties are included. The data are shown at different values of $x$, as indicated on the plot.}
\label{fig:cc}
\end{center}
\end{figure}

We also perform the fit with $Q^2_{\rm min}=3.5~\GeV^2$ in order 
to compare in detail with the results in \cite{Abramowicz:2015mha}, where 
this is their default cut. In Table~\ref{tab:t1} we show the breakdown of 
$\chi^2$ values for the different HERA neutral and charged current data 
sets. We include the numbers for the global fit including the HERA combined 
data, as well as the results for  the fit to the HERA data only, at both NLO 
and NNLO. There appears 
to be some tension between the $e^- p$ charged current data and other data in 
the global fit, with the NLO fit to the HERA only data giving a $\chi^2$ for 
these data which 
is $\sim$ 20 units higher than the global fits. The tension is somewhat lower at 
NNLO, where the increase is $\sim$ 10 units less. 
The $\chi^2$ for the neutral current data at $920$ GeV also shows some, 
albeit relatively lower, sensitivity to whether a global fit is performed.

In Fig.~\ref{fig:cc} we show the data/theory at NNLO for the $e^-$ 
charged current 
data in 
different $x$ bins. It can be seen that while the local fit gives a good 
description of the data, the comparison for the global fit has a different 
shape. It tends to largely overshoot the data at intermediate $x$, i.e. 
in bins $x=0.032, 0.08, 0.13$, but generally undershoots it at higher $x$. 
These charged current data are mainly sensitive to the up (at high $x$ valence) 
quark. Hence, in the global fit data other than HERA data, in practice 
largely fixed proton target DIS data, clearly prefer a  
different shape for the up quark.
In particular, the HERA charged current data prefers a somewhat smaller/larger $u$ quark at 
intermediate/larger $x$ compared to the other global data.  
We will return to this in the next section.

\section{Effect on the PDFs}\label{sec:pdfs}

\begin{figure}[h]
\begin{center}
\includegraphics[scale=0.66]{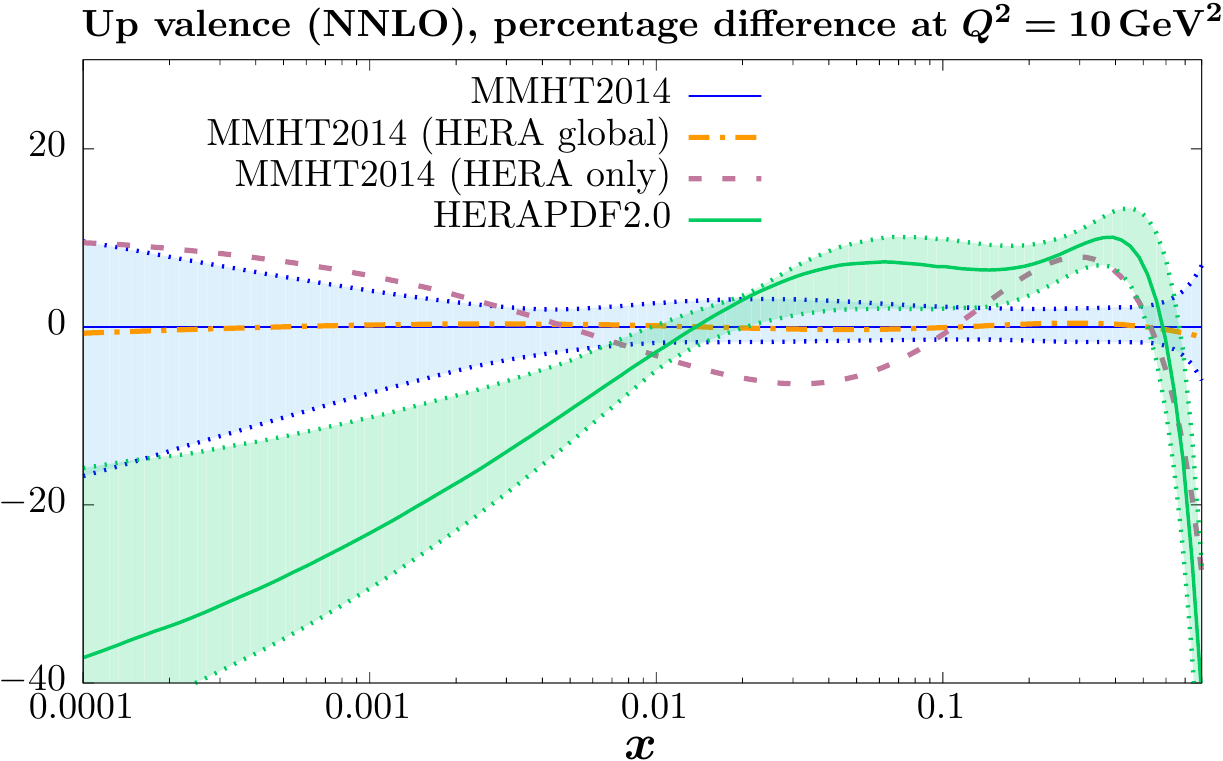}\!\!
\includegraphics[scale=0.66]{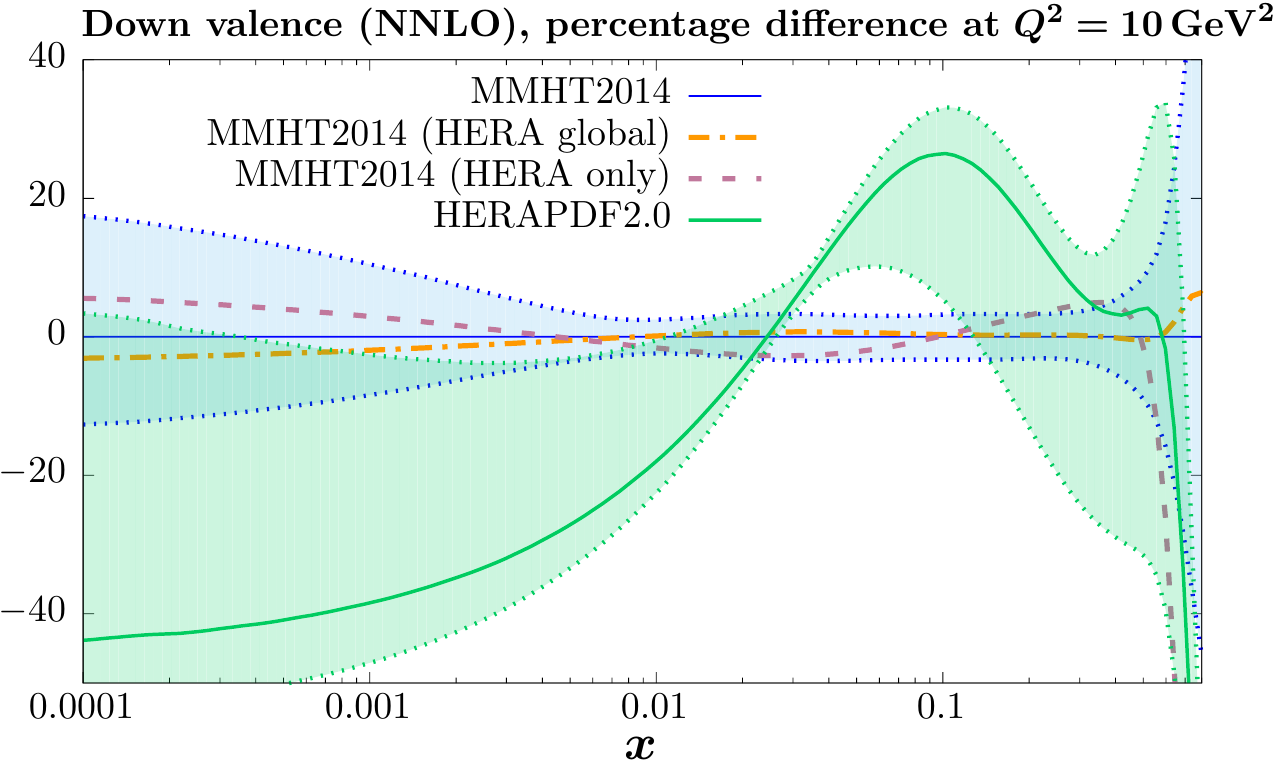}
\!\!\!\!\!\!\!\!\!
\includegraphics[scale=0.66]{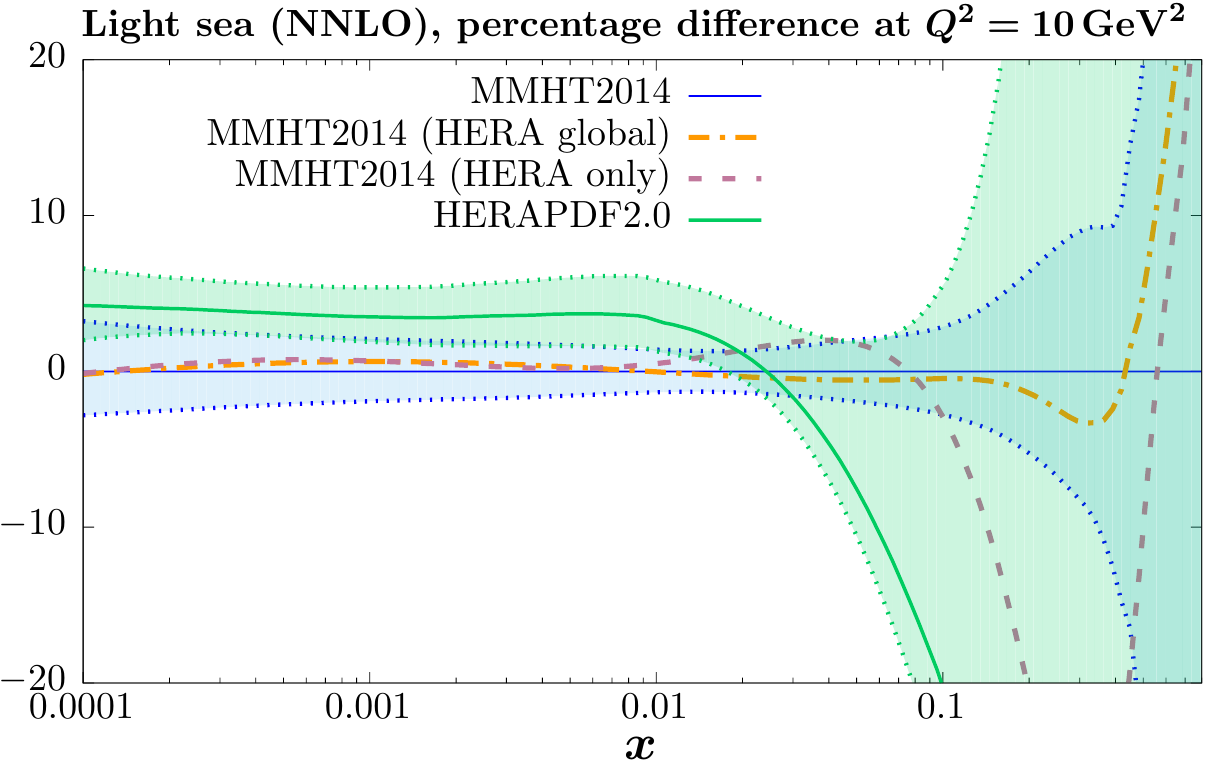}
\includegraphics[scale=0.66]{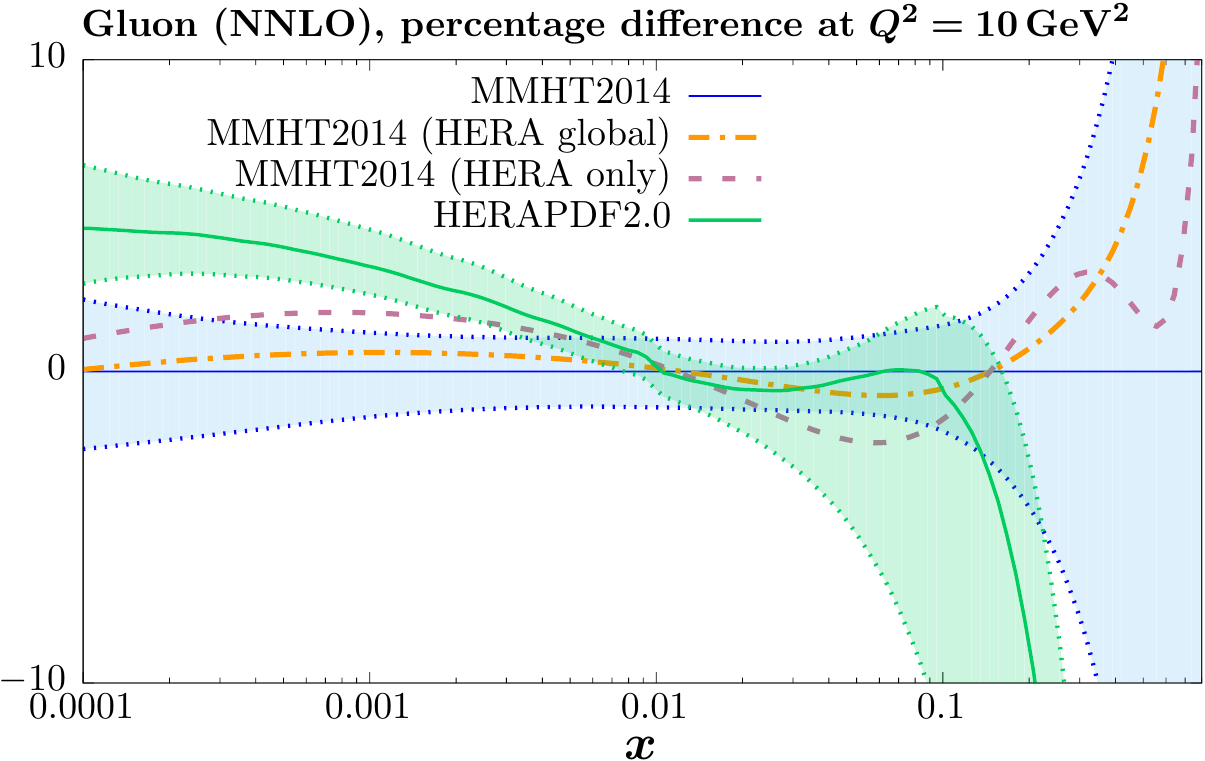}
\caption{\sf Comparison between the up and down valence, gluon and light 
quark sea distributions at $Q^2=10^4~\GeV^2$ for the standard MMHT2014 fit, 
with the corresponding PDF uncertainties, with the central values of the fit including the HERA combined 
data, as well as the fit to only this data set, shown as dot--dashed and dashed curves, respectively. Also shown are the 
HERAPDF2.0 distributions, including PDF uncertainties.}
\label{fig:PDFs}
\end{center}
\end{figure}

Since the fit quality does not improve very significantly from the prediction
using the MMHT 2014 PDFs we do not expect much change in the central value of 
the PDFs in the new global fit which includes the HERA I+II combined data. More change might be expected in the PDFs fit to
only HERA data as then the main constraints on some types of PDF are lost. 
In Fig.~\ref{fig:PDFs} we show the central values of the NNLO PDFs from the 
fits
including the new HERA combined data, comparing them to MMHT2014 PDFs (with 
uncertainties) and the HERAPDF2.0 PDFs (also with uncertainties). The modified 
global PDFs are always very well within the MMHT2014  
uncertainty bands.

The PDFs from the fit to only HERA run I $+$ II data are 
in some ways similar to those of HERAPDF2.0, e.g. the up valence 
quark for $x>0.2$, which shows some significant 
deviations from the global fits PDF set. This appears to be driven 
by the $e^-$ charged current data, but there is clearly tension with the 
rest of the data in the global fit, as our full fit including the new HERA 
data does not have this feature. Similarly, the sea quarks in our fit to only 
HERA data prefer to be soft at high $x$, like for HERAPDF2.0, but in this case
there is no real constraint on high-$x$ sea quarks from HERA DIS data, and the 
HERAPDF2.0 uncertainty band is not in conflict with the global fits.  
However, the common features between our fit to only HERA run I $+$ II data
and HERAPDF2.0 are not 
universal -- the gluon and the down valence distributions in our fit to 
only HERA data are much more similar to 
MMHT2014 than HERAPDF2.0. This is likely to be a feature of the differing 
parameterisations used in the two studies. The very high-$x$ gluon in 
the global fits definitely prefers a harder gluon than in HERAPDF2.0, due to 
constraints from jet data and fixed target DIS data, but even in our HERA data
only fit, there is no actual preference for the softer high-$x$ gluon. 
Also, we certainly see no suggestion of 
HERA data preferring a significantly different shape down valence 
distribution to that preferred by other sets in the global fit, and our 
central value in the HERA data only fit is surprisingly close to that in 
our global fits given the relative lack of constraint on this distribution 
from HERA DIS data.

\begin{figure}[h]
\begin{center}
\includegraphics[scale=0.66]{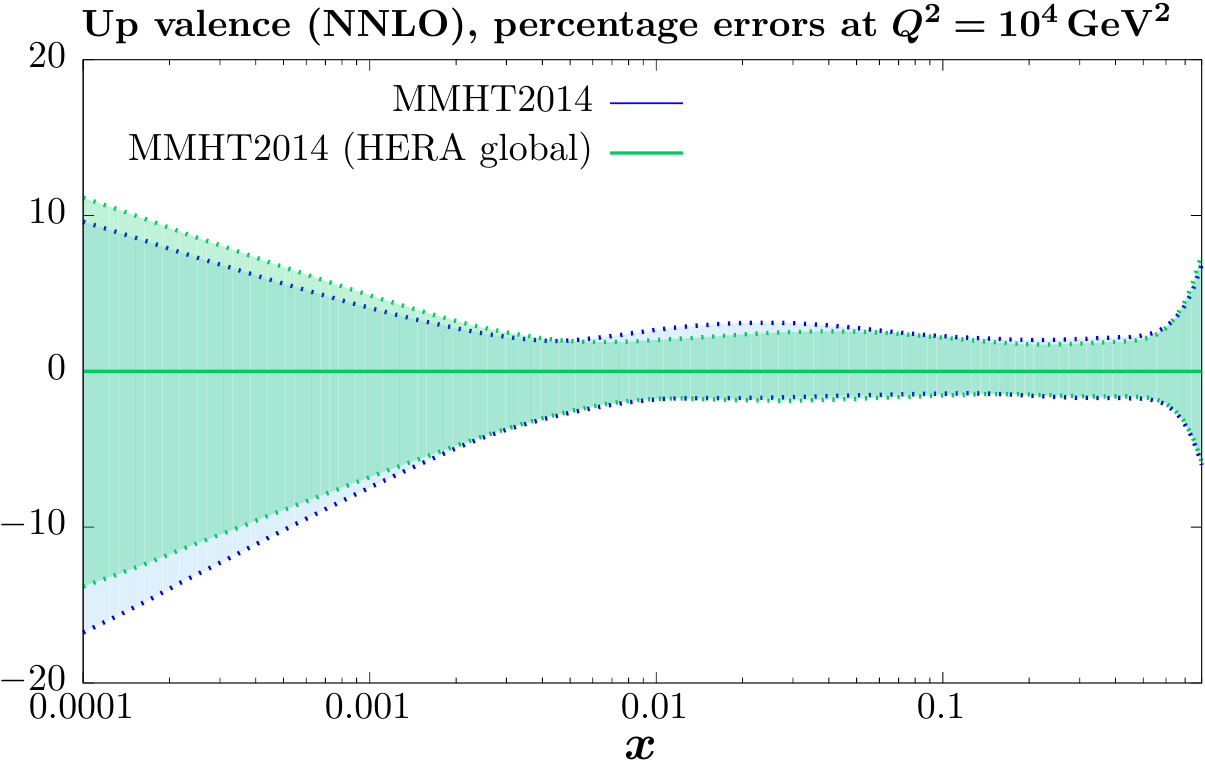}
\includegraphics[scale=0.66]{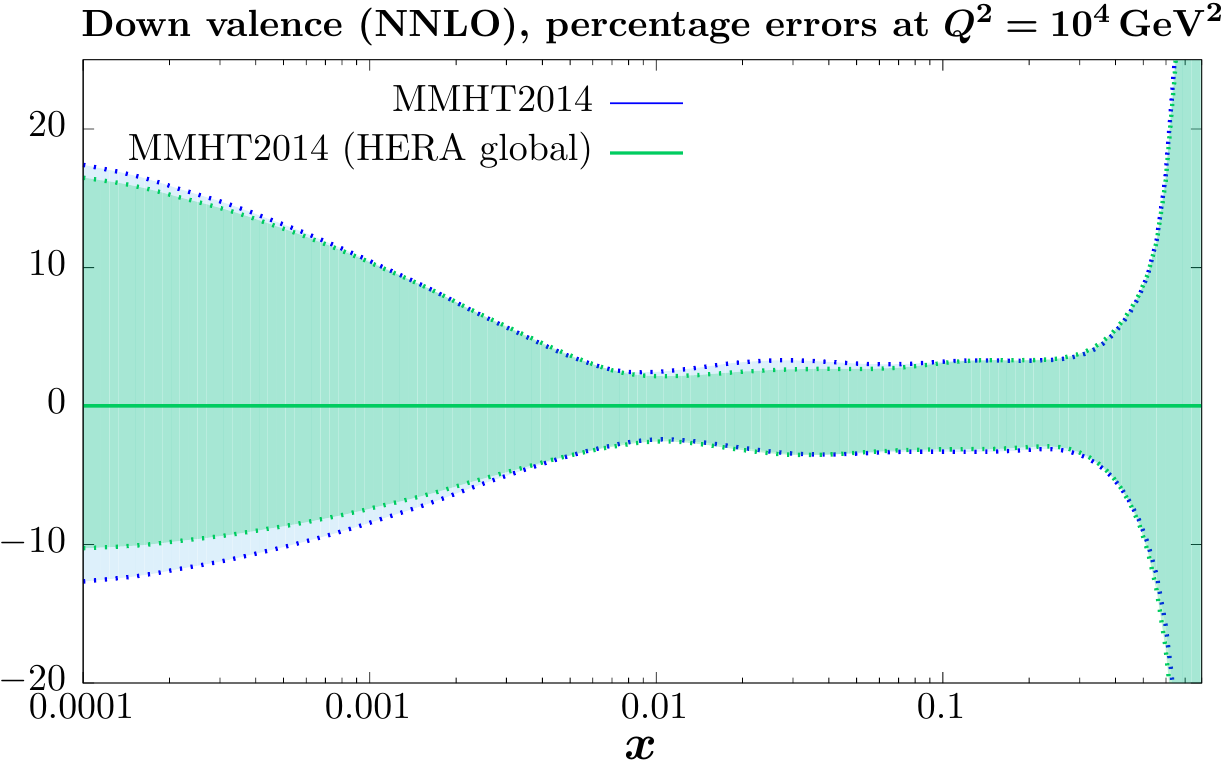}
\includegraphics[scale=0.66]{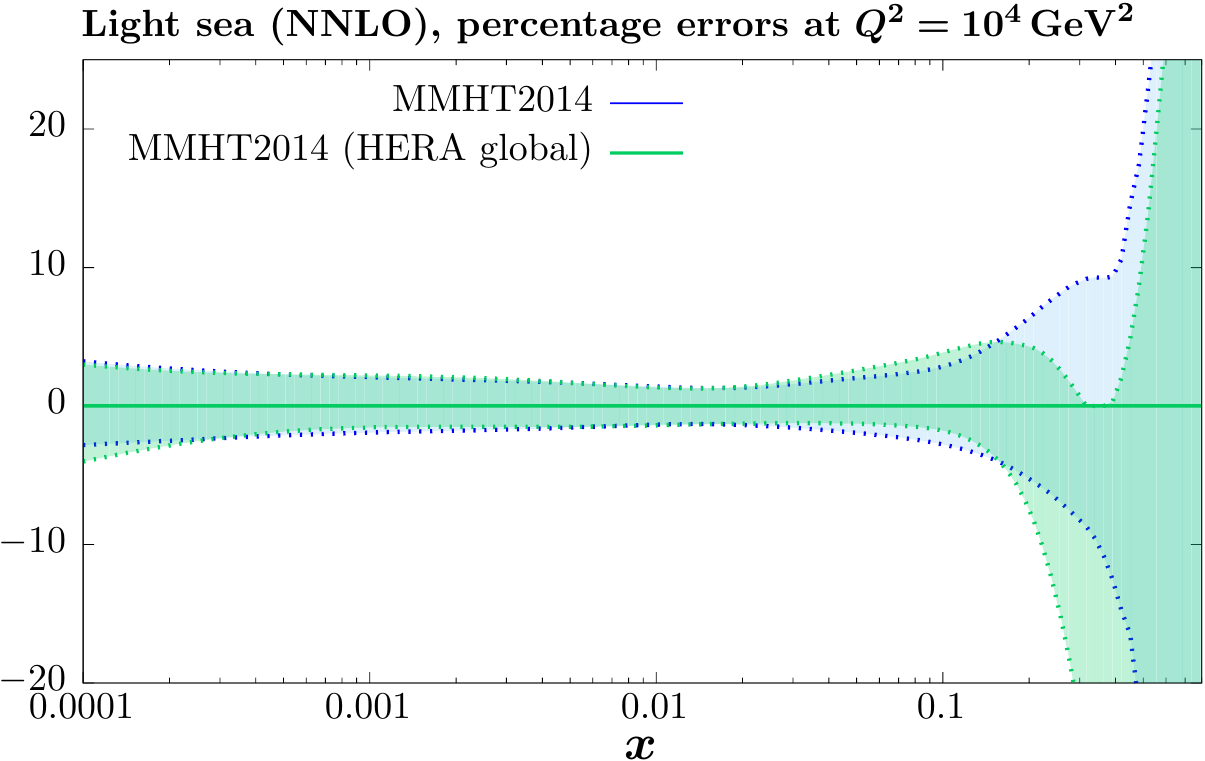}
\includegraphics[scale=0.66]{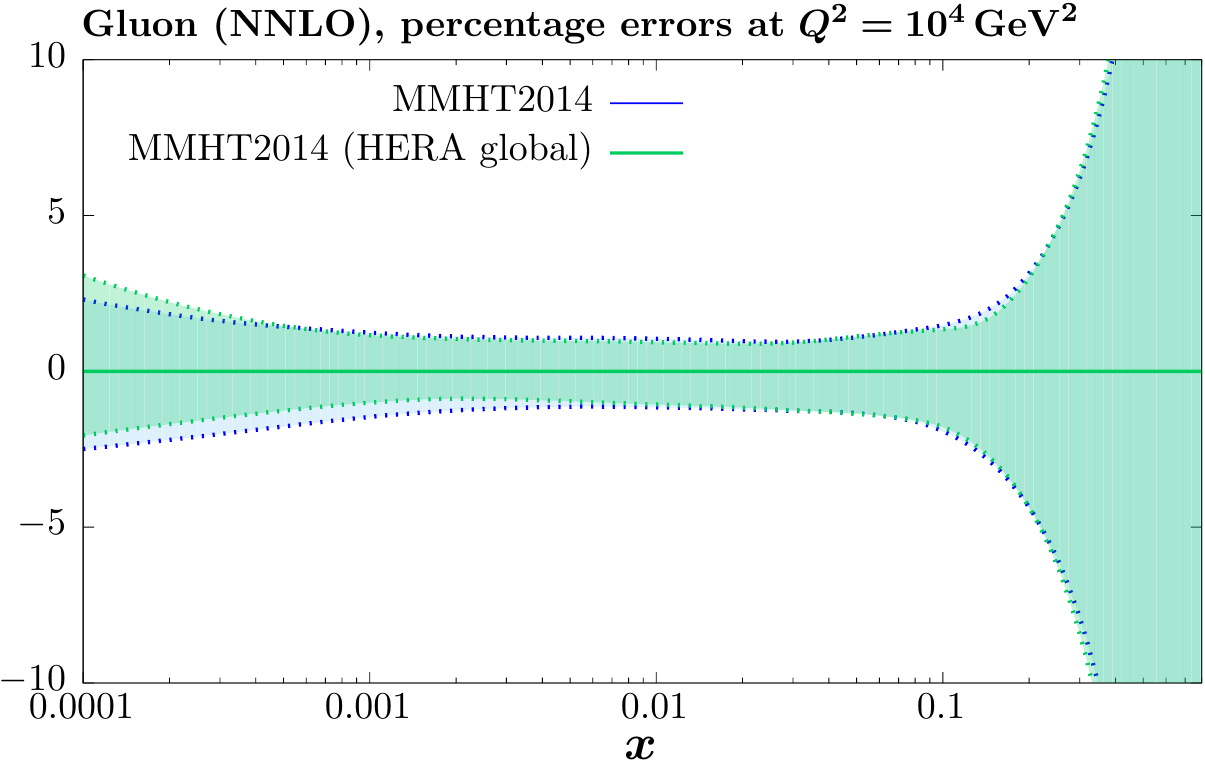}
\caption{\sf Comparison between the up and down valence, gluon and light 
quark sea distributions at $Q^2=10^4~\GeV^2$ for the MMHT2014 set and 
the corresponding uncertainties and the fit 
including the HERA combined data set with their corresponding uncertainties. }
\label{fig:PDFserr}
\end{center}
\end{figure}

We also investigate the effect of the new HERA data on the uncertainties of
the PDFs. In order to determine PDF uncertainties we use the 
same ``dynamic tolerance'' 
prescription to determine eigenvectors as for MSTW2008 \cite{MSTW}.
In Fig.~\ref{fig:PDFserr} we compare the uncertainties for the NNLO PDFs
including the HERA run I $+$ II data in a global fit to the uncertainties of 
the MMHT2014 
PDFs. These are very similar to MMHT2014 in most features.
The most obvious improvement from the inclusion of the new HERA data 
is to the gluon for $x < 0.01$. 
There is also a slight improvement in some places for the valence quarks, but 
the additional constraint supplied by much improved charged current data is 
overwhelmed by the constraint of valence quark PDFs from other data in the 
global fit. 
While the improvements generally appear to be quite moderate, in fact when 
benchmark cross section predictions are considered, the effect of the HERA 
combined data in reducing the corresponding PDF uncertainties becomes 
somewhat clearer; we consider this in the following section.

\section{Effect on benchmark cross sections}\label{sec:benchmark}

In Table~\ref{tab:sigmaNNLO} we show NNLO predictions for benchmark $W,Z$, 
Higgs and $t\overline{t}$ cross sections at a range of collider energies, 
for the standard MMHT14 PDF set, and for the result of the same fit, but 
including the HERA combined data.

To calculate the cross section we use the same procedure as was used 
in~\cite{Harland-Lang:2014zoa}. That is, for $W, Z$ and Higgs production we 
use the code provided by W.J. Stirling, based on the
calculation in~\cite{WZNNLO},~\cite{HiggsNNLO1} and~\cite{HiggsNNLO2}, and 
for top pair production
we use the procedure and code of \cite{topNNLO}. Here our primary aim is not 
to present definitive predictions or to compare in detail to other PDF sets, 
as both these results are frequently provided in the literature with very 
specific choices of codes, scales and parameters which may differ from those 
used here.  Rather, our main objective is to illustrate the effect that 
the combined HERA data has on the central values and uncertainties of the 
cross sections.

\begin{table}
\begin{center}
\vspace{-1.0cm}
\begin{tabular}{|l|c|c|}
\hline
& MMHT14 & MMHT14 (HERA global)   \\
\hline
$\!\! W\,\, {\rm Tevatron}\,\,(1.96~\TeV)$   &$2.782^{+0.056}_{-0.056}$ $\left({}^{+2.0\%}_{-2.0\%}\right)$& $2.789^{+0.050}_{-0.050}$  $\left({}^{+1.8\%}_{-1.8\%}\right)$ \\   
$\!\! Z \,\,{\rm Tevatron}\,\,(1.96~\TeV)$  & $0.2559^{+0.0052}_{-0.0046}$ $\left({}^{+2.0\%}_{-1.8\%}\right)$ &$0.2563^{+0.0047}_{-0.0047}$  $\left({}^{+1.8\%}_{-1.8\%}\right)$\\    
$\!\! W^+ \,\,{\rm LHC}\,\, (7~\TeV)$        &$6.197^{+0.103}_{-0.092}$  $\left({}^{+1.7\%}_{-1.5\%}\right)$ & $6.221^{+0.100}_{-0.096}$ $\left({}^{+1.6\%}_{-1.5\%}\right)$ \\    
$\!\! W^- \,\,{\rm LHC}\,\, (7~\TeV)$        & $4.306^{+0.067}_{-0.076}$  $\left({}^{+1.6\%}_{-1.8\%}\right)$&$4.320^{+0.064}_{-0.070}$ $\left({}^{+1.5\%}_{-1.6\%}\right)$  \\    
$\!\! Z \,\,{\rm LHC}\,\, (7~\TeV)$          & $0.964^{+0.014}_{-0.013}$  $\left({}^{+1.5\%}_{-1.3\%}\right)$& $0.966^{+0.015}_{-0.013}$ $\left({}^{+1.6\%}_{-1.3\%}\right)$   \\    
$\!\! W^+ \,\,{\rm LHC}\,\, (14~\TeV)$       & $12.48^{+0.22}_{-0.18}$ $\left({}^{+1.8\%}_{-1.4\%}\right)$& $12.52^{+0.22}_{-0.18}$ $\left({}^{+1.8\%}_{-1.4\%}\right)$ \\    
$\!\! W^- \,\,{\rm LHC}\,\, (14~\TeV)$      & $9.32^{+0.15}_{-0.14}$ $\left({}^{+1.6\%}_{-1.5\%}\right)$ & $9.36^{+0.14}_{-0.13}$  $\left({}^{+1.5\%}_{-1.4\%}\right)$  \\    
$\!\! Z \,\,{\rm LHC}\,\, (14~\TeV)$       & $2.065^{+0.035}_{-0.030}$ $\left({}^{+1.7\%}_{-1.5\%}\right)$&   $2.073^{+0.036}_{-0.026}$ $\left({}^{+1.7\%}_{-1.3\%}\right)$ \\    
\hline    
$\!\! {\rm Higgs} \,\,{\rm Tevatron}$       & $0.874^{+0.024}_{-0.030}$ $\left({}^{+2.7\%}_{-3.4\%}\right)$&$0.866^{+0.019}_{-0.023}$   $\left({}^{+2.2\%}_{-2.7\%}\right)$   \\
$\!\!{\rm Higgs} \,\,{\rm LHC}\,\,(7~\TeV)$ & $14.56^{+0.21}_{-0.29}$   $\left({}^{+1.4\%}_{-2.0\%}\right)$ &$14.52^{+0.19}_{-0.24}$  $\left({}^{+1.3\%}_{-1.7\%}\right)$ \\
$\!\!{\rm Higgs} \,\,{\rm LHC}\,\,(14~\TeV)$ & $47.69^{+0.63}_{-0.88}$  $\left({}^{+1.3\%}_{-1.8\%}\right)$ & $47.75^{+0.59}_{-0.72}$ $\left({}^{+1.2\%}_{-1.5\%}\right)$ \\ 
\hline    
$\!\! t\bar t \,\,{\rm Tevatron}$             & $7.51^{+0.21}_{-0.20}$ $\left({}^{+2.8\%}_{-2.7\%}\right)$&    $7.57^{+0.18}_{-0.18}$ $\left({}^{+2.4\%}_{-2.4\%}\right)$\\
$\!\! t\bar t\,\,{\rm LHC}\,\,(7~\TeV)$     & $175.9^{+3.9}_{-5.5}$   $\left({}^{+2.2\%}_{-3.1\%}\right)$&  $174.8^{+3.3}_{-5.3}$   $\left({}^{+1.9\%}_{-3.0\%}\right)$ \\
$\!\! t\bar t\,\,{\rm LHC}\,\,(14~\TeV)$      & $970^{+16}_{-20}$ $\left({}^{+1.6\%}_{-2.1\%}\right)$&$964^{+13}_{-19}$   $\left({}^{+1.3\%}_{-2.0\%}\right)$ \\ 

\hline
    \end{tabular}
\end{center}
\caption{\sf The values of various cross sections (in nb) obtained with the NNLO MMHT 2014 sets, with and without the final HERA combination data set included.  PDF uncertainties only are shown.}
\label{tab:sigmaNNLO}   
\end{table}

For $W,Z$ production the central values of the predicted cross sections are 
only slightly affected by the inclusion of the HERA data, while there is some 
small, i.e. up to a few $\%$ level, reduction in the PDF uncertainties. 
For Higgs Boson production the predicted cross sections again change very 
little - well within PDF uncertainties. However, here the reduction in PDF 
uncertainty is larger, up to $\sim 10 \%$ of the MMHT uncertainty. Finally, 
for $t\overline{t}$ production the picture is similar to the Higgs case, with 
the central value relatively unchanged, and the uncertainties reduced at 
the $\sim 10 \%$ level. This highlights that the new HERA data provides 
some extra constraint within the global fit, but mainly due to the reduced uncertainty
on the gluon 
distribution for the LHC predictions. 

\section{Investigation of $Q^2_{\rm min}$ dependence}\label{sec:q2min}

\begin{figure}
\begin{center}
\vspace*{-1.0cm}
\includegraphics[scale=0.41]{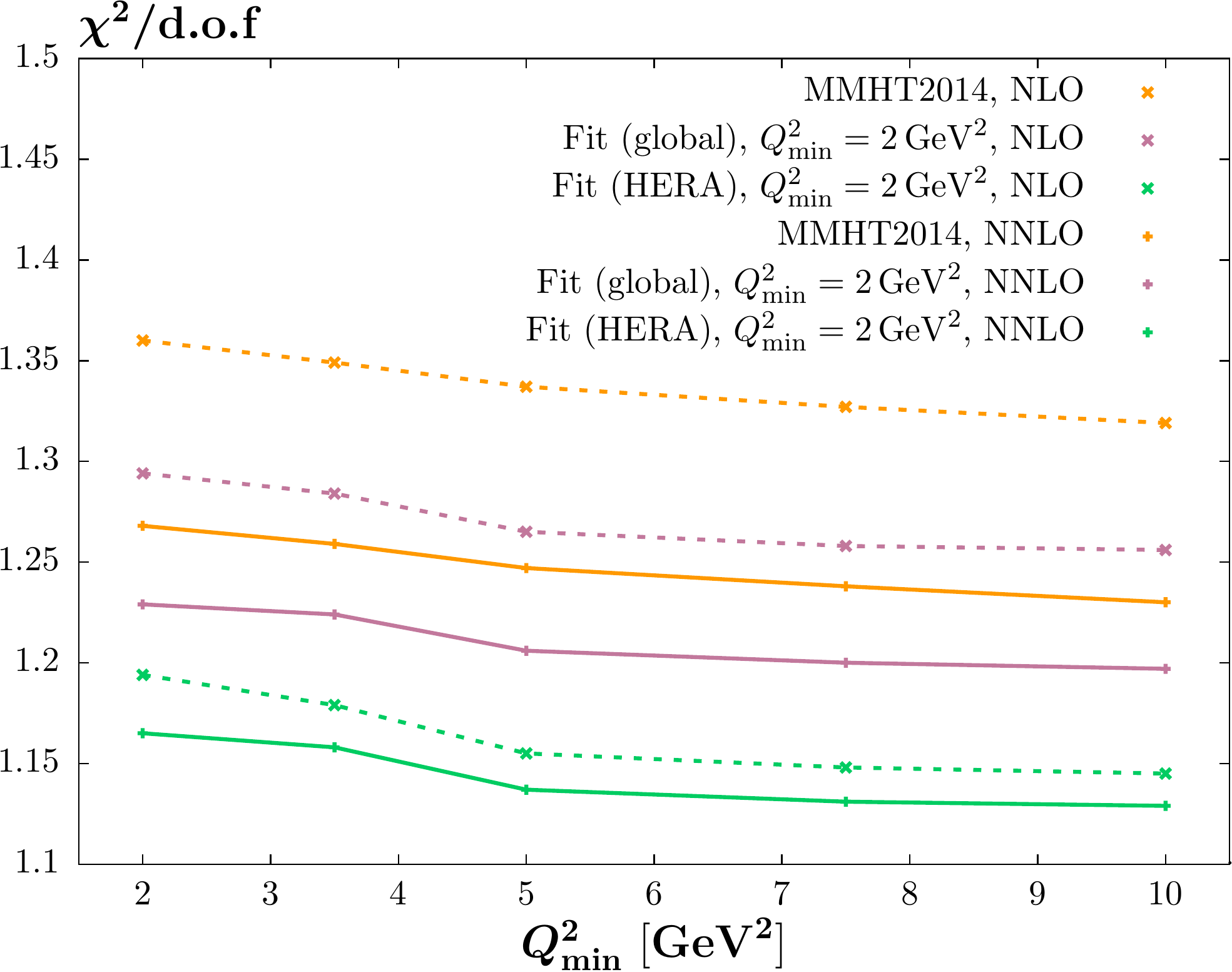}
\caption{\sf The $\chi^2$ per degree of freedom for the MMHT2014 
{\it predictions} (which occur in the plot in descending order) to the HERA combined 
data set, 
and for the global + HERA combined 
and HERA combined only fits, with $Q^2_{\rm min}=2\,{\rm GeV}^2$ fixed; the plot 
versus $Q^2_{\rm min}$ is then obtained by calculating the $\chi^2$/d.o.f. for the HERA
combined data with $Q^2>Q^2_{\min}$. The NLO (NNLO) curves are shown as dashed 
(continuous) curves.} 
\label{fig:chicomp1}
\end{center}
\end{figure}

The HERAPDF2.0 analysis sees a marked improvement in $\chi^2$ per point with 
a raising of the $Q^2_{\min}$ value for the data fit. Hence, we also 
investigate the variation of the fit quality for changes of $Q^2_{\min}$. 
However, to begin with 
we simply calculate the quality of the comparison to data as a function of 
$Q^2_{\min}$ at NLO and at NNLO without performing a refit, 
i.e. the PDFs used were those obtained
with the default $Q^2_{\min}=2~\GeV^2$ cut. This is shown in Fig.~\ref{fig:chicomp1}
where we show a comparison of the $\chi^2$ per point for 
the three variations of NLO and NNLO comparisons, i.e. the 
MMHT2014 prediction, 
the global refit including the new HERA data and the refit with 
only HERA run I $+$ II
combined data. From the figure it is clear that NNLO is always 
superior, but this is less 
distinct in the refits, particularly for the fit to 
only HERA data. It is also clear there is a reasonable 
lowering of the $\chi^2$ per point as $Q^2_{\min}$ increases, but no 
clear ``jumps'' in improvement. 

\begin{figure}
\begin{center}
\vspace*{-1.0cm}
\includegraphics[scale=0.41]{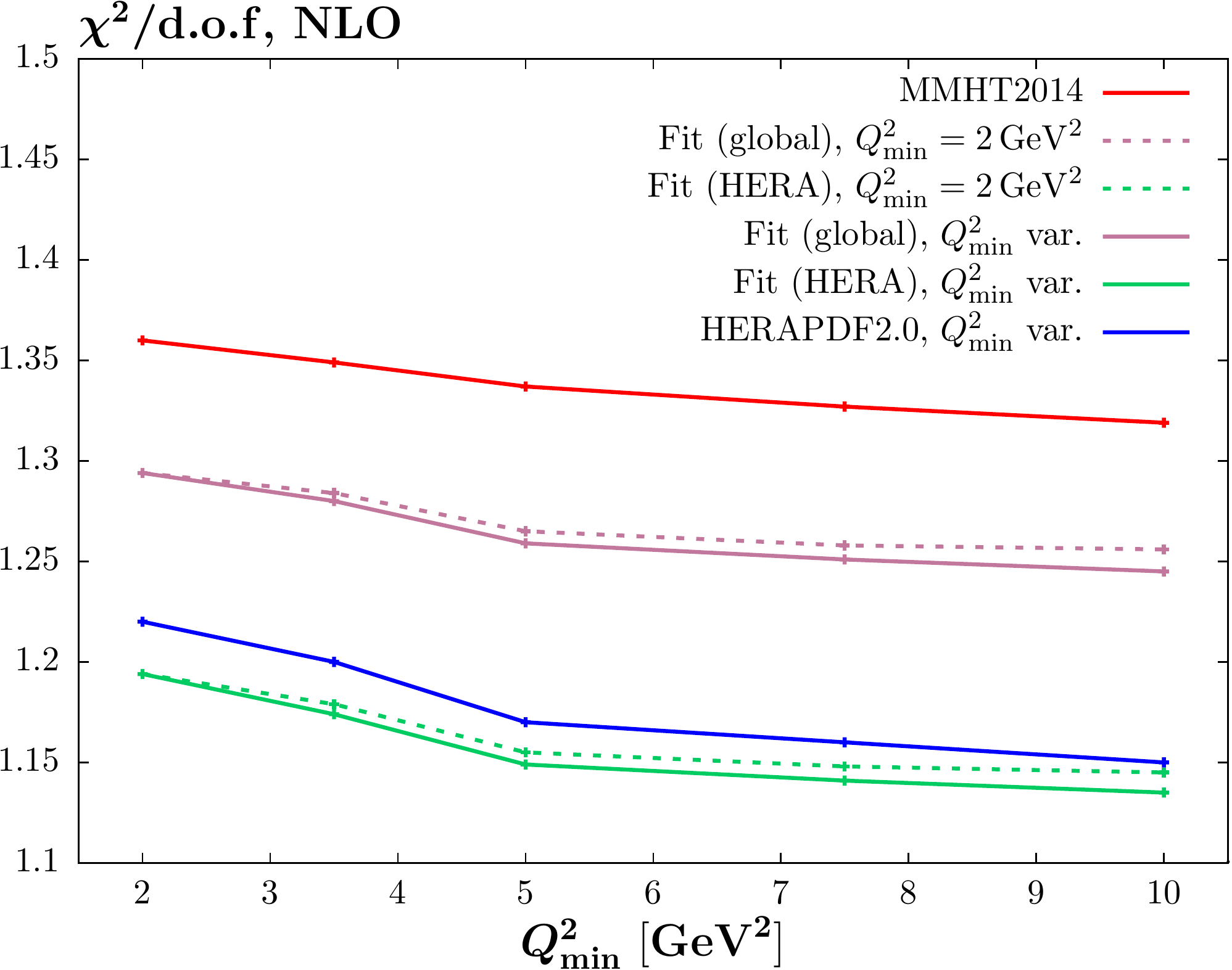}
\includegraphics[scale=0.41]{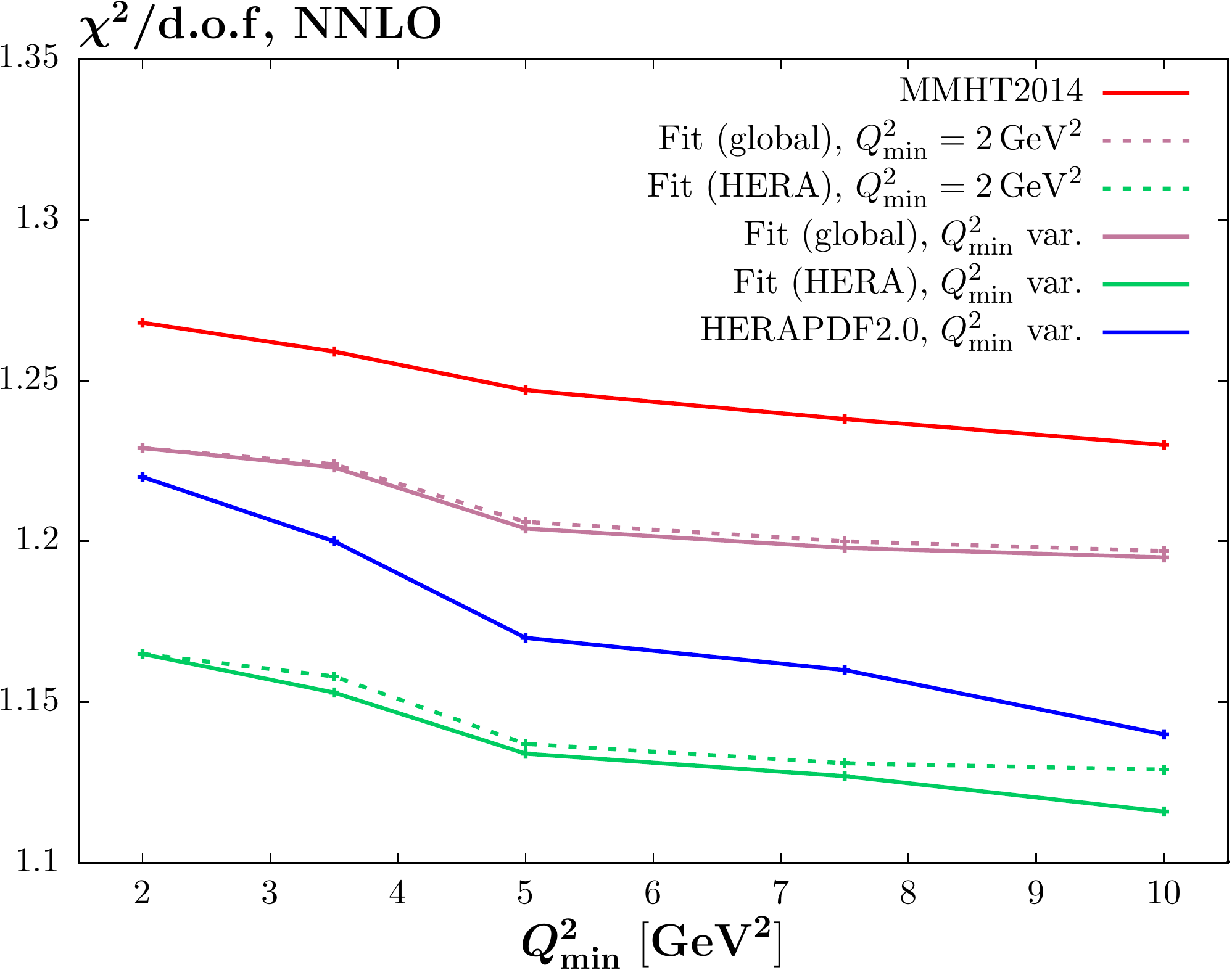}
\caption{\sf The $\chi^2$ per degree of freedom for the MMHT2014 {\it predictions} to the
HERA combined data set, and for the global + HERA combined and HERA combined only 
fits, with $Q^2_{\rm min}=2\,{\rm GeV}^2$; the plot versus $Q^2_{\rm min}$ is then
obtained by calculating the $\chi^2$ contribution from the HERA combined data with
$Q^2>Q^2_{\rm min}$. These are shown (reproduced from 
Fig.~\ref{fig:chicomp1}) as dashed curves, while the two solid curves just below these  show the effect of
{\it fits} with $Q^2_{\min}$ varied (rather than fixed at $Q^2_{\min}=2~\GeV^2$).
The result of the  HERAPDF2.0 fit with varying $Q^2_{\rm min}$ is also shown.
The left/right hand figure shows the NLO/NNLO fits.}
\label{fig:chicomp2}
\end{center}
\end{figure}

We also look at the effect of changing the $Q^2$ cut in the fit itself 
(though we change the cut only for the HERA combined data, not for the 
 other data in the global fit), 
at both NLO and NNLO. This is shown in Fig.~\ref{fig:chicomp2}, where we 
also show 
the trend for the HERAPDF2.0 analysis \cite{Abramowicz:2015mha}.\footnote{The 
definition of $\chi^2$ for the HERAPDF2.0 fit is not identical. However, 
this should be a very small effect.} For comparison we also include the 
curves from Fig.~\ref{fig:chicomp1} for the 
$\chi^2$ per point obtained for varying $Q^2_{\min}$ but with the 
fits performed for $Q^2_{\min}=2~\GeV^2$.  
We note that while there is an improvement 
in $\chi^2$ per point with increasing $Q^2_{\min}$, as observed in 
\cite{Abramowicz:2015mha}, this is very largely
achieved without any refitting. This is more marked in the global fit, where
(at NNLO in particular) the refit with raised $Q^2_{\min}$ has only a 
minimal effect. It is very clear there is also less improvement with 
$Q^2_{\min}$ in our 
analysis than for HERAPDF2.0, particularly in the 
global fit and at NNLO. This may be due to our more extensive PDF 
parameterisation obtaining shapes that manage to fit the lowest $Q^2$ data
better.

\section{Effect of higher-twist type corrections}
\begin{figure}
\begin{center}
\vspace*{-1.0cm}
\includegraphics[scale=0.41]{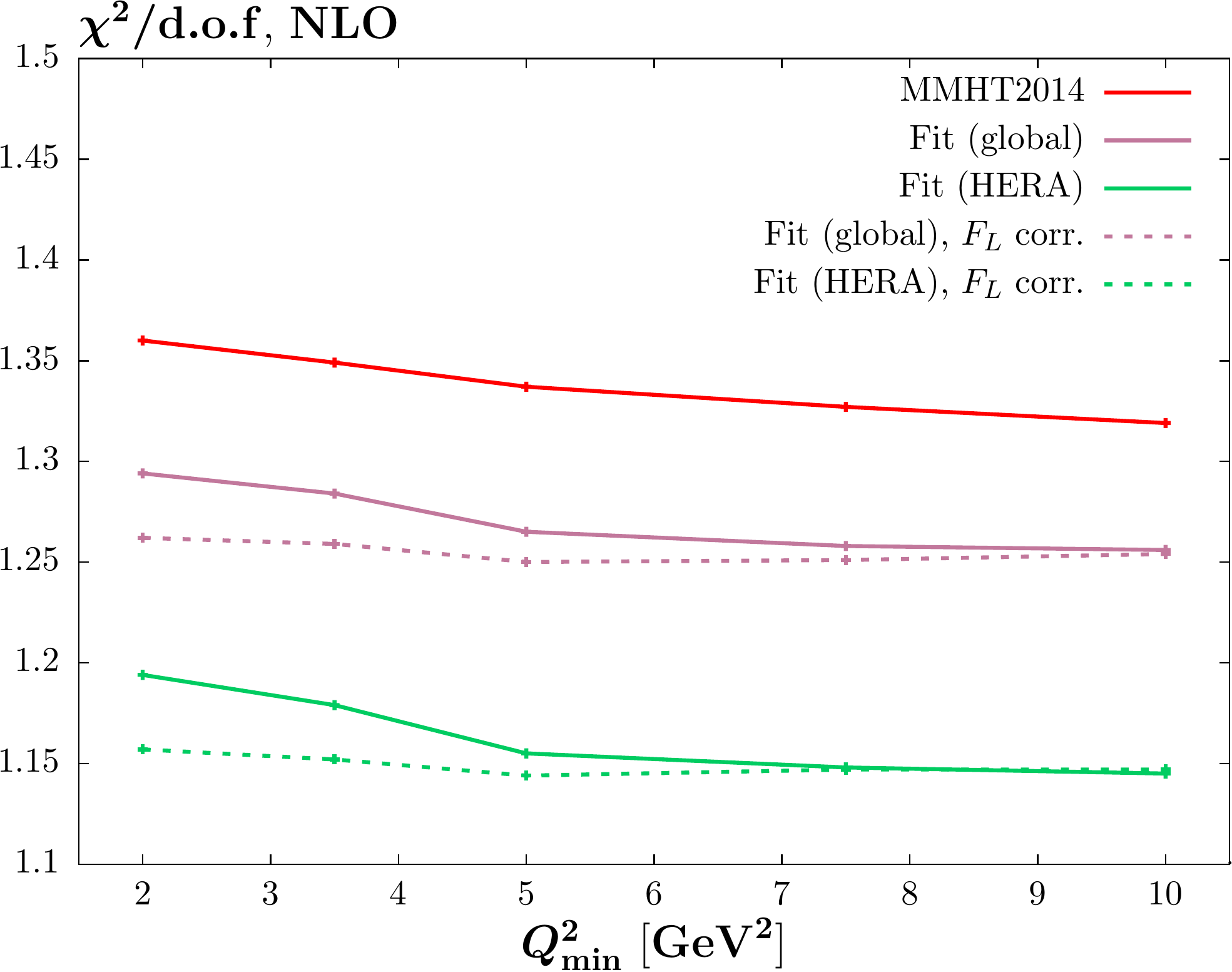}
\includegraphics[scale=0.60]{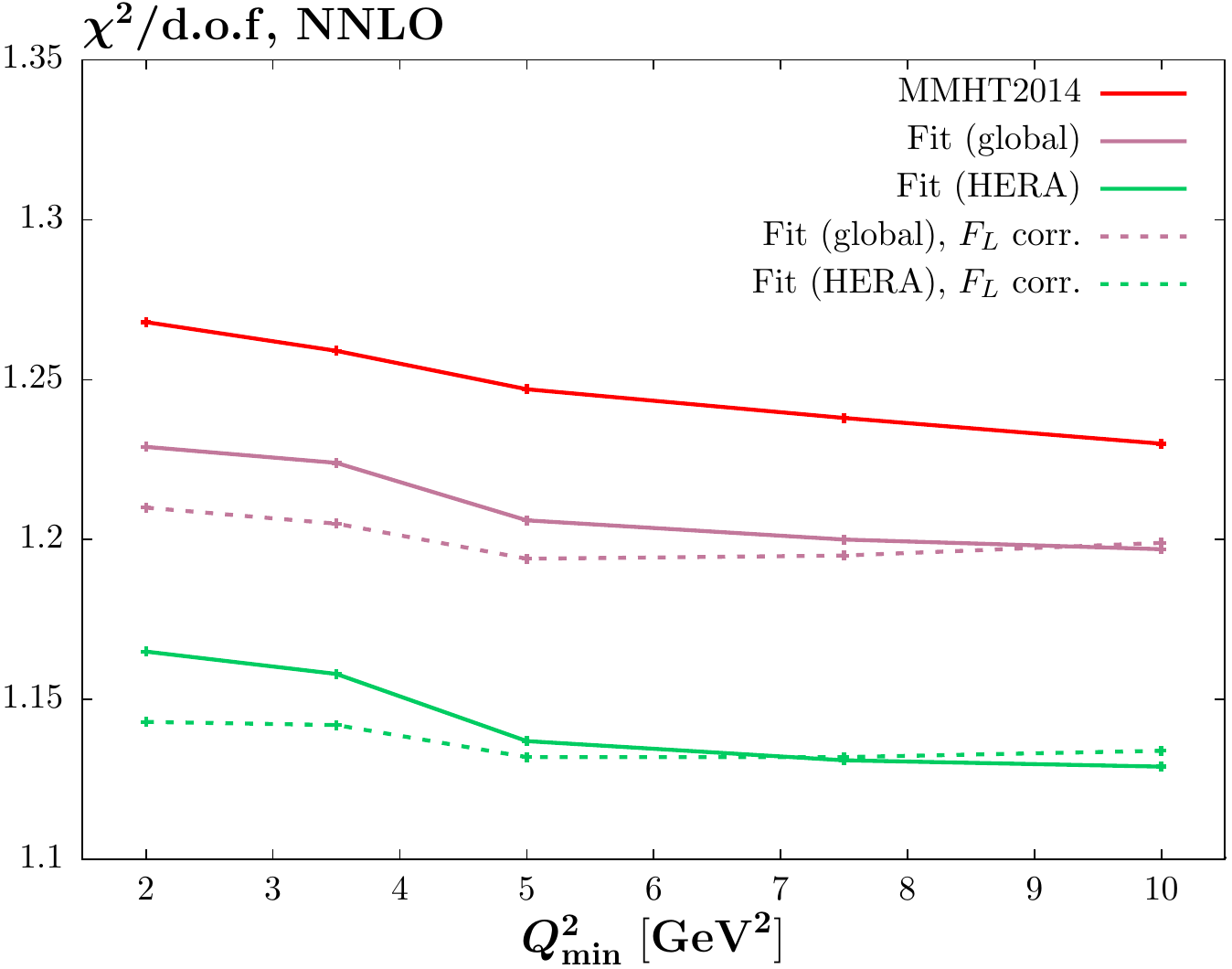}
\caption{\sf The behaviour of the $\chi^2$ per degree of 
freedom when we include the higher twist correction (\ref{eq:flcorr}), shown by the dashed curves, as compared
to the curves of Fig.~\ref{fig:chicomp1} which were obtained without the correction. 
The left/right hand figure shows the NLO/NNLO fits. 
}
\label{fig:chicomp3}
\end{center}
\end{figure}

In order to investigate the possibility of improving the $\chi^2$ per point for low
$Q^2_{\min}$ we will consider some simple phenomenological corrections to the 
reduced cross section
\begin{equation}
\tilde{\sigma}(x,Q^2)=F_2(x,Q^2)-\frac{y^2}{1+(1-y^2)}F_L(x,Q^2)\;.
\end{equation}
As much of the deterioration in fit quality with decreasing $Q^2_{\rm min}$ 
seems to occur due to a general tendency of the fit to overshoot the HERA 
neutral current data at highest $y$ and low $x$ and $Q^2$, the region where 
the $F_L$ contribution is most important, we will first 
consider corrections to the $F_L$ theory prediction, before commenting on 
$F_2$.
Motivated by the possible contribution of higher twist corrections, we  
consider the very simple possibility
\begin{equation}\label{eq:flcorr}
F_L^{(1)}(x,Q^2)=F_L(x,Q^2)\left(1+\frac{a}{Q^2}\right)\;.
\end{equation}
Allowing the parameter $a$ to be free and performing a refit, we find a 
reduction in $\Delta \chi^2=24$ in the default 
($Q^2_{\rm min}=2\,{\rm GeV}^2$) NNLO fit (and very similar at NLO), with 
quite a large value of $a=4.30\,{\rm GeV}^2$. As this correction will be 
concentrated in the lower 
$Q^2$ region we may expect this to affect the trend observed in 
Figs.~\ref{fig:chicomp1} and~\ref{fig:chicomp2} with $Q^2_{\rm min}$. 
In Fig.~\ref{fig:chicomp3} we show the $\chi^2/{\rm dof}$ with (\ref{eq:flcorr}) applied by the dashed curves, and compare with the curves of Fig.~\ref{fig:chicomp1}. The effect is 
significant, flattening the behaviour essentially entirely. We notice, 
however, that for the highest $Q^2_{\min}$ considered, i.e. 
$Q^2_{\min}=10~\GeV^2$, the $\chi^2$ obtained with the PDFs and $F_L$ 
corrections for $Q^2_{\min}=2~\GeV^2$ can be marginally higher than 
for the fits obtained for $Q^2_{\min}=2~\GeV^2$ without the $F_L$ 
correction. It we perform a refit for each value of $Q^2_{\min}$ then, as 
in Section 5 the improvement in fit quality is minimal, but this feature for
$Q^2_{\min}=10\,\GeV^2$ is removed, and for this higher cut the preferred
$F_L$ correction is smaller.

\begin{figure}
\begin{center}
\vspace*{-1.0cm}
\includegraphics[scale=0.4]{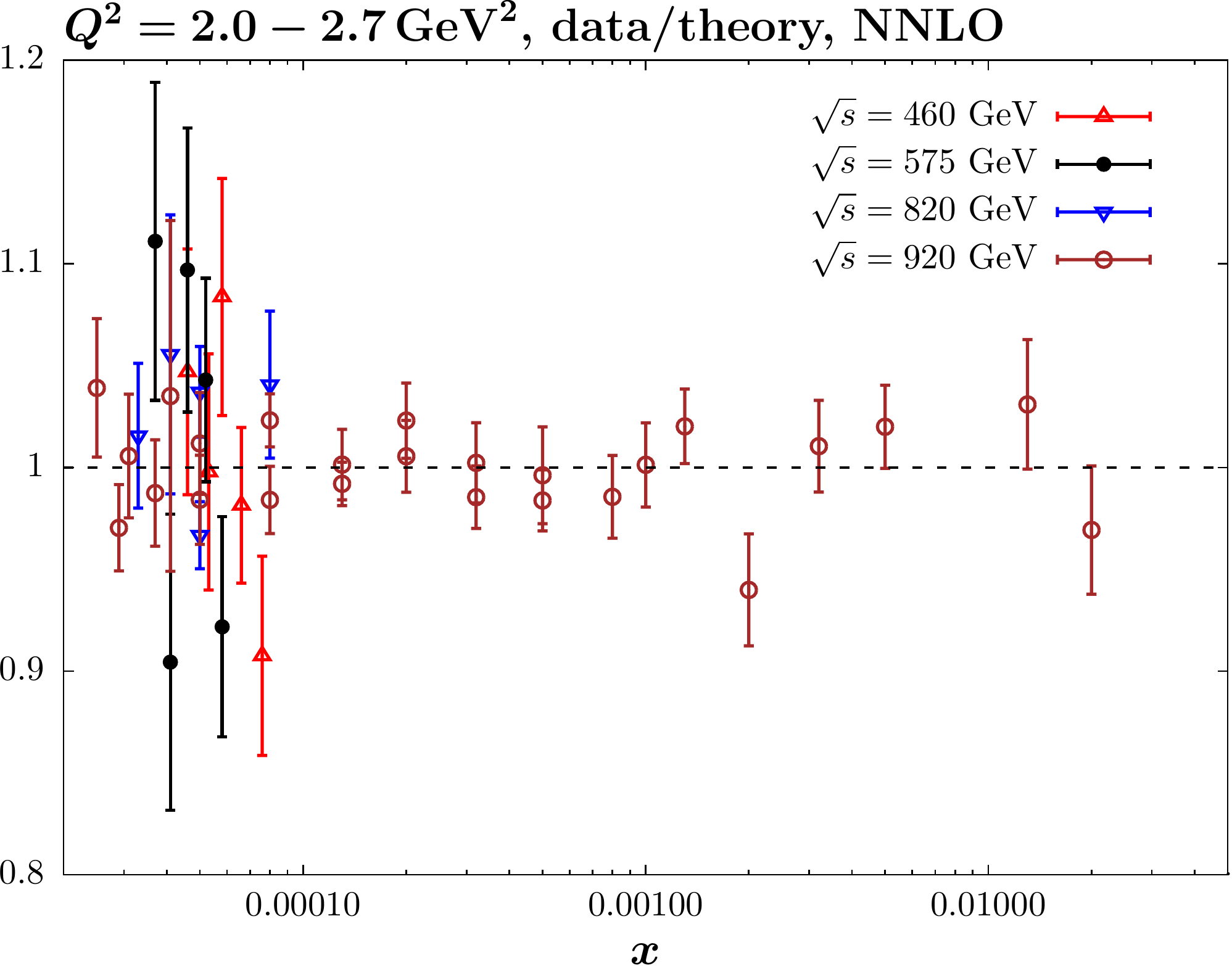}
\includegraphics[scale=0.4]{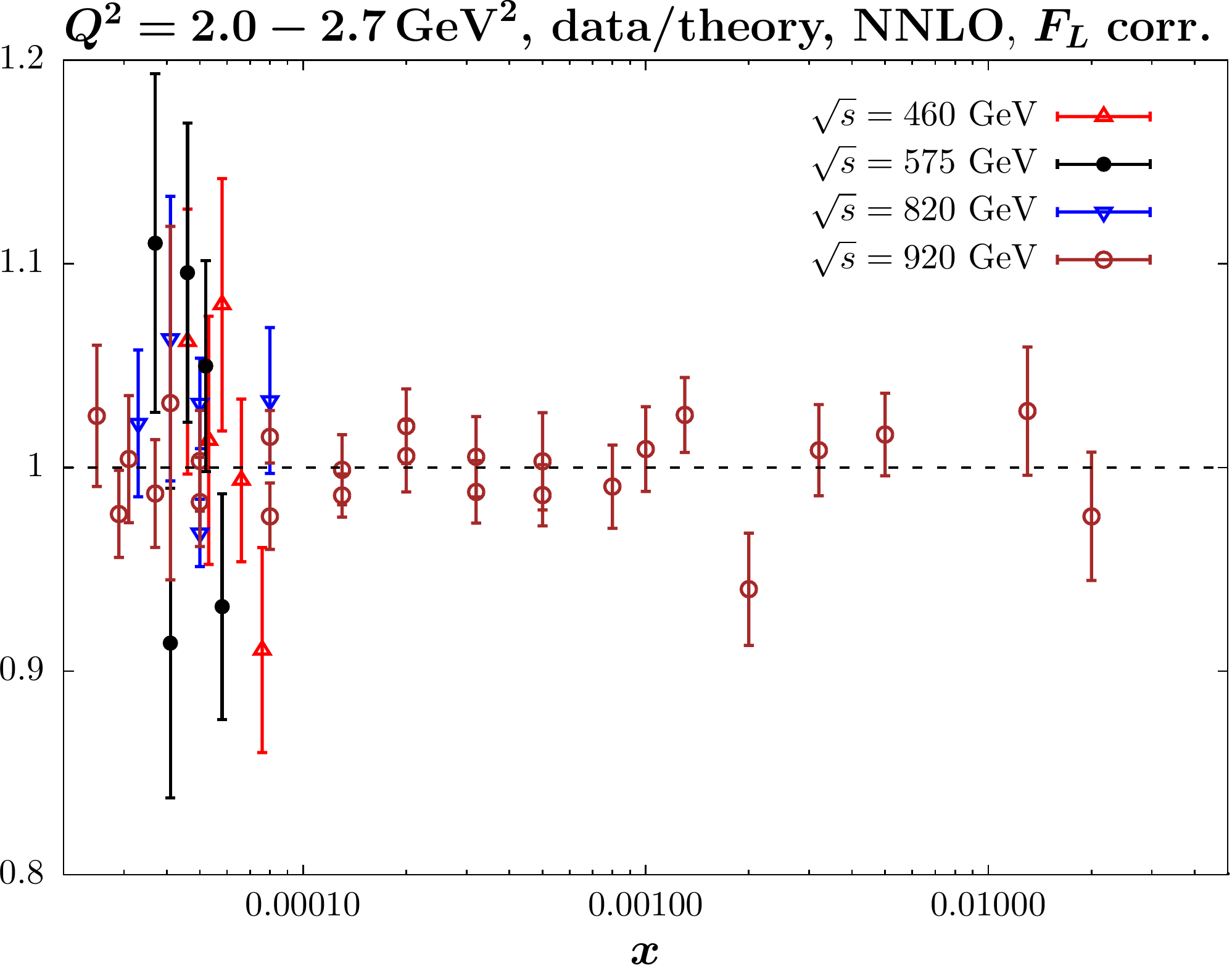}
\includegraphics[scale=0.4]{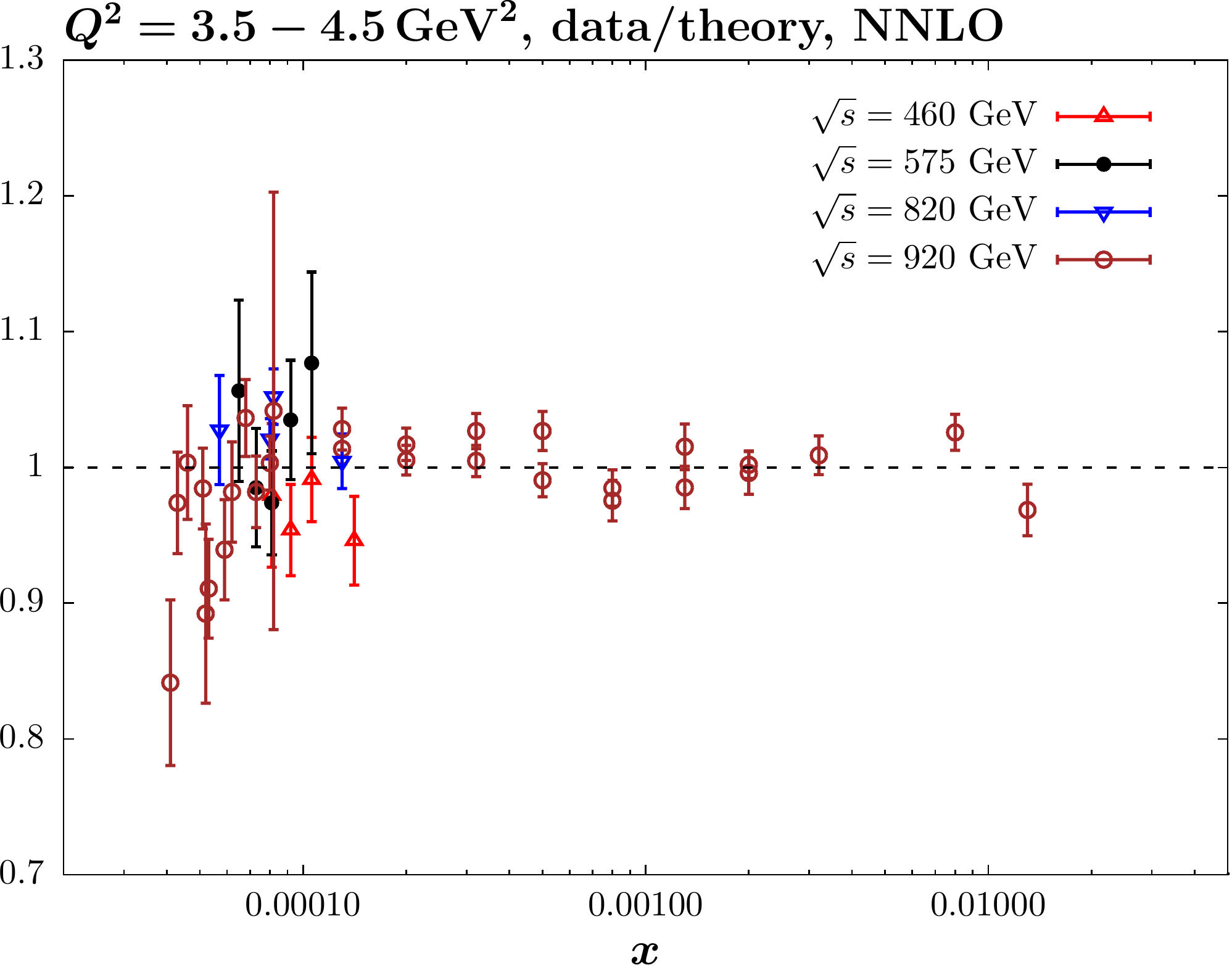}
\includegraphics[scale=0.4]{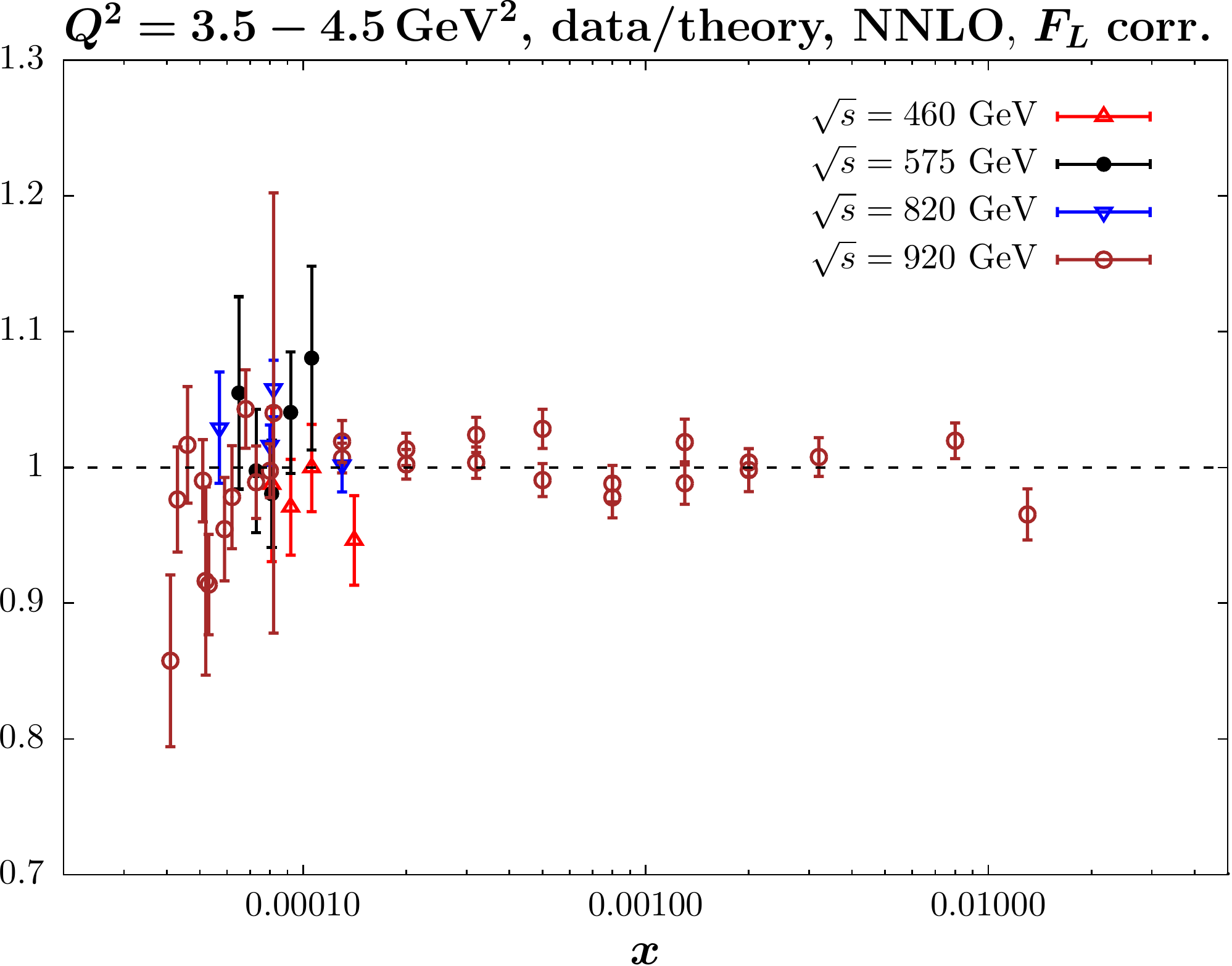}
\includegraphics[scale=0.4]{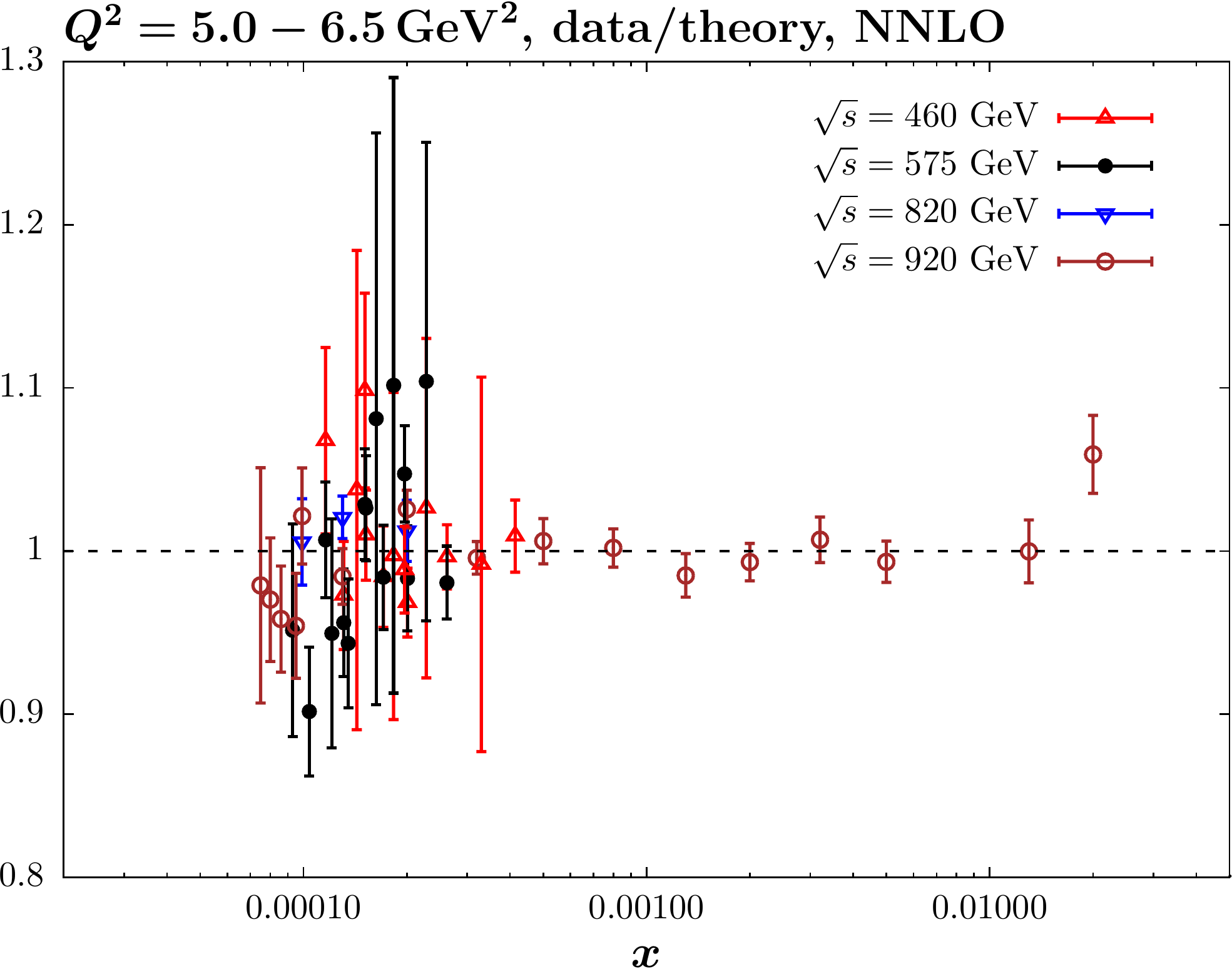}
\includegraphics[scale=0.4]{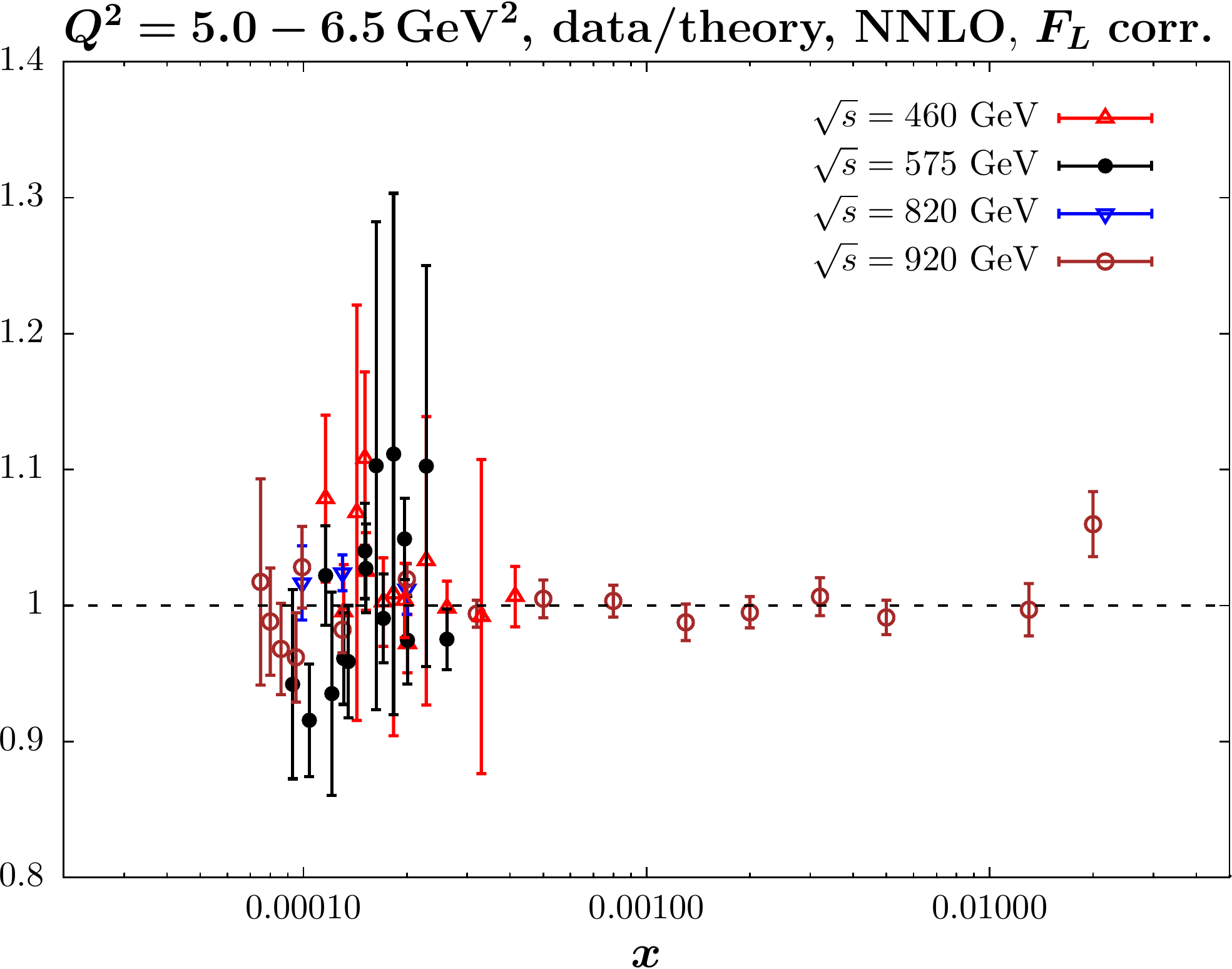}
\caption{\sf HERA NC data/theory for global MMHT fit including HERA combined data without (left) and with (right) the correction (\ref{eq:flcorr}) applied, divided into individual data sets and for three ranges of $Q^2=2.0-2.7,\, 3.5-4.5,\, 5.0-6.5\,{\rm GeV}^2$. The shifts of data relative to theory due to correlated uncertainties are included.
}
\label{fig:fldata}
\end{center}
\end{figure}

\begin{figure}
\begin{center}
\vspace*{-1.0cm}
\includegraphics[scale=0.6]{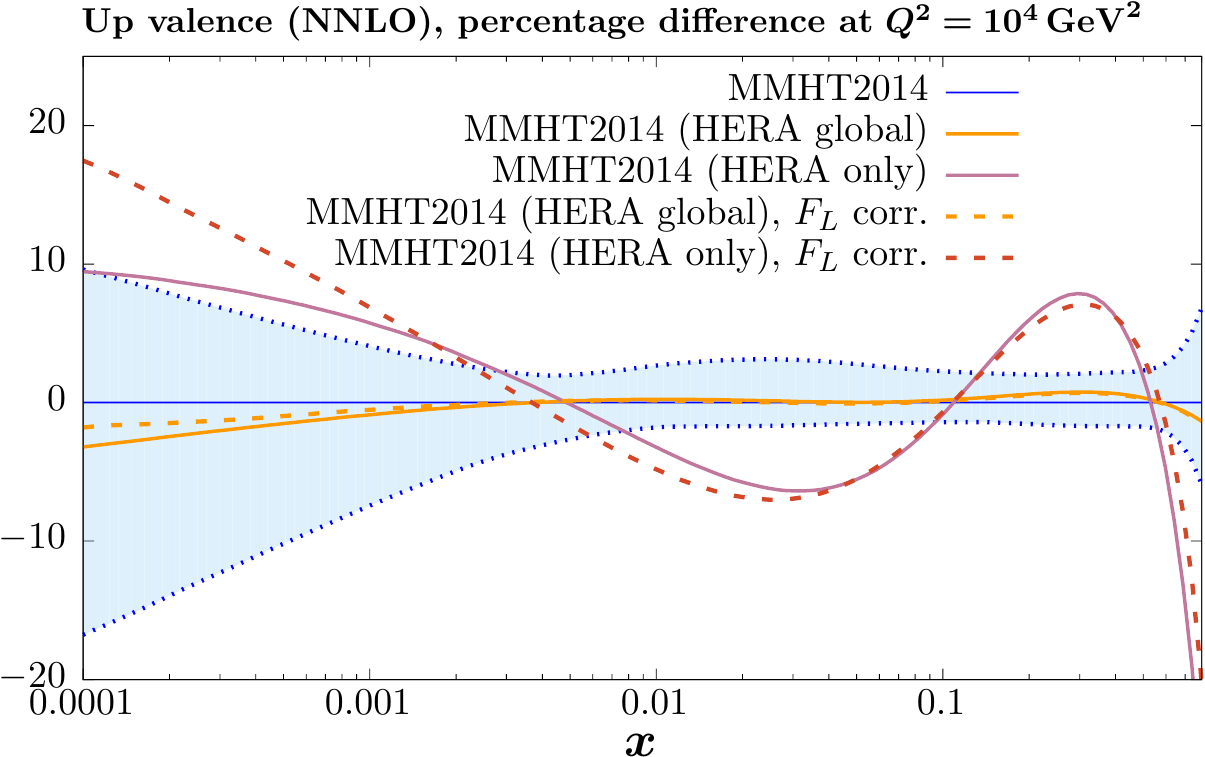}
\includegraphics[scale=0.6]{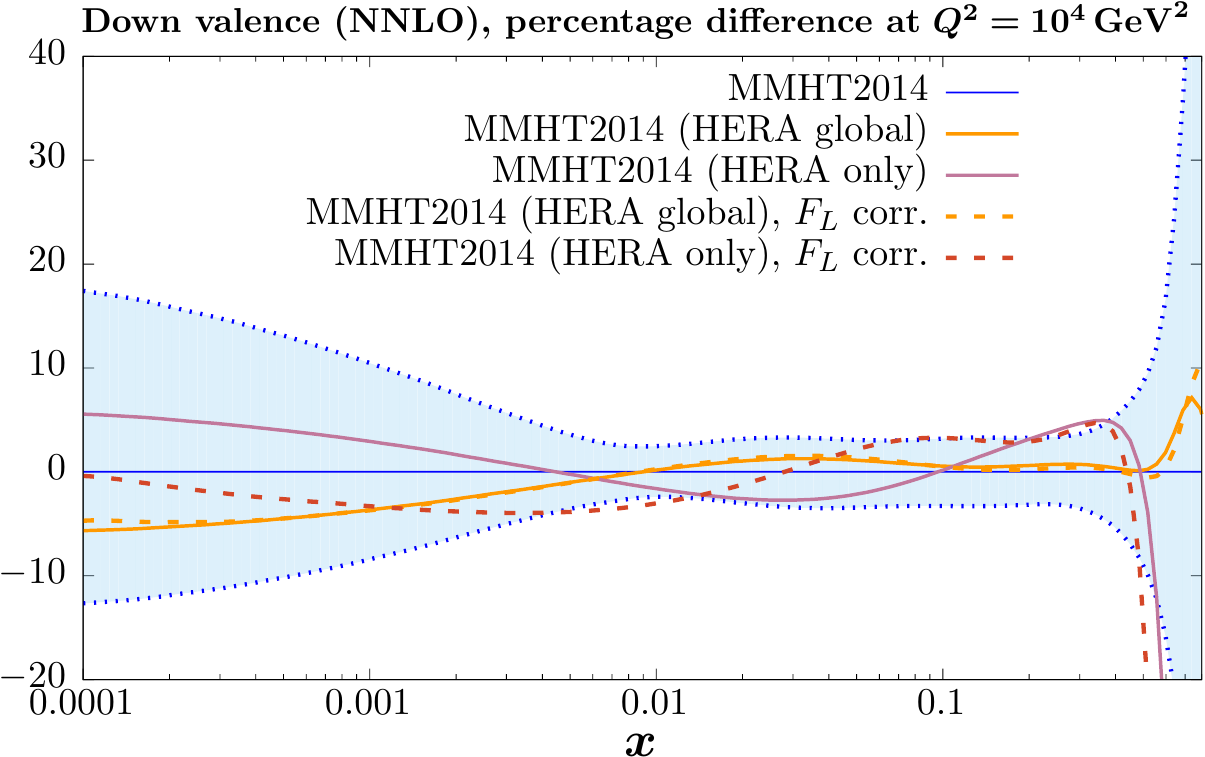}
\includegraphics[scale=0.6]{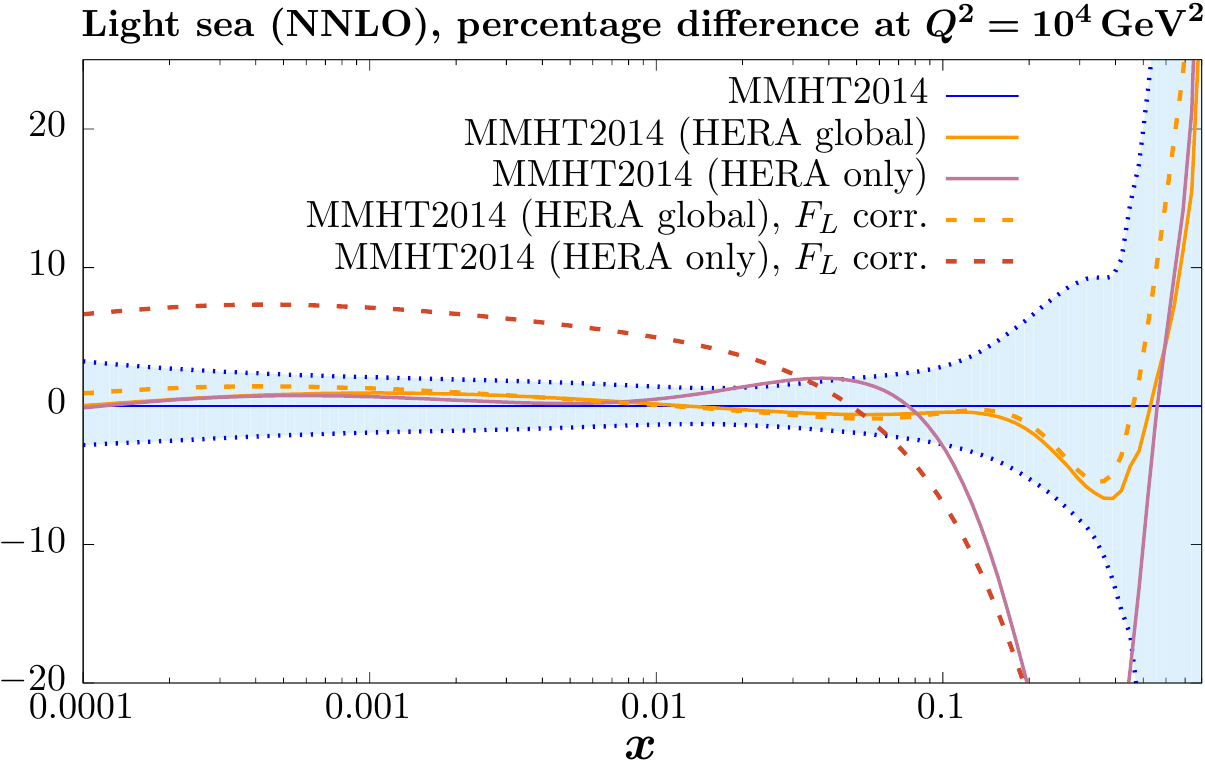}
\includegraphics[scale=0.6]{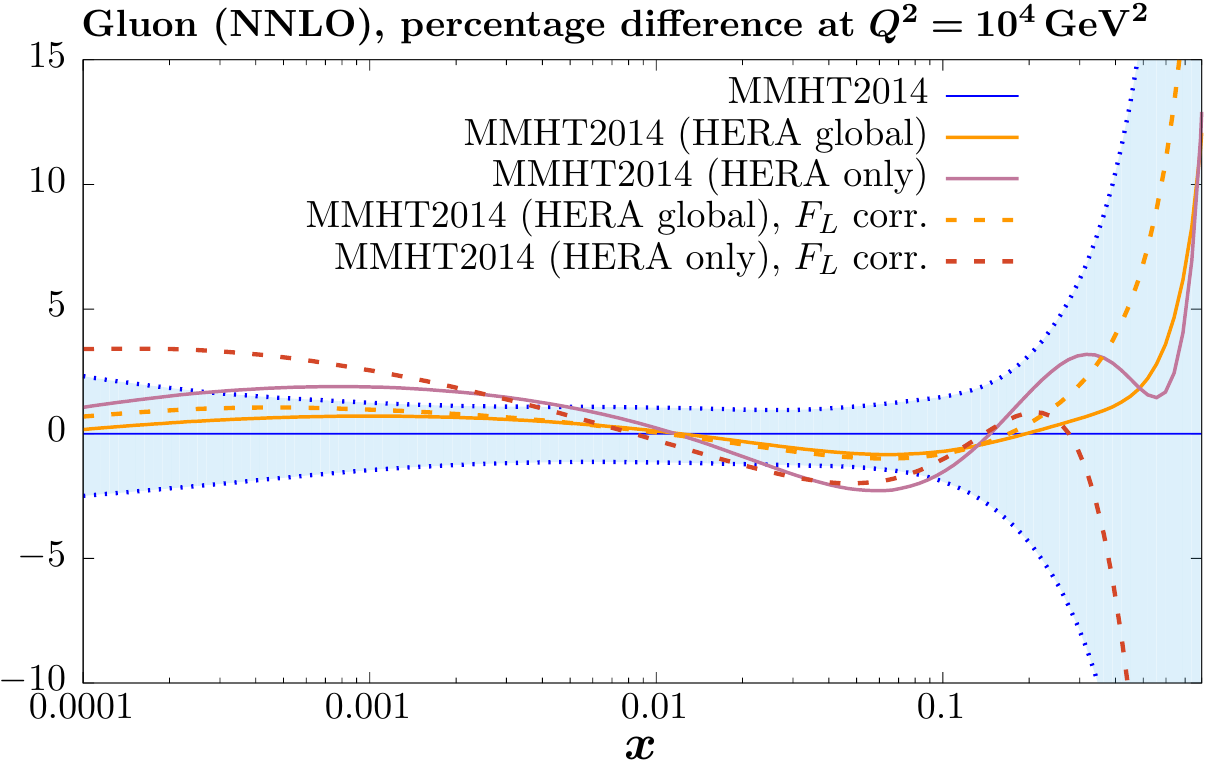}
\caption{\sf Comparison between the up and down valence, gluon and light quark sea distributions at $Q^2=10^4~\GeV^2$ for the standard MMHT2014 fit, with the 
MMHT2014 PDF errors, and for the central fits including the HERA combined data, as well as the fit to only this data set, with and without the correction (\ref{eq:flcorr}) applied to $F_L$.}
\label{fig:PDFsfl}
\end{center}
\end{figure}

To get a clearer 
picture, we can look at the effect on the neutral current data/theory comparison. 
This is shown in Fig.~\ref{fig:fldata} with and without this correction 
applied. As seen in the left-hand plots there is a tendency to overshoot
some of the highest $y$ points, and while this is not eliminated entirely
for all points by the correction, some tightening of the data/theory is 
evident and the scatter is more consistent with fluctuations . 
It is worth pointing out that some of the improvement in $\chi^2$ actually 
comes from a reduction in the shift in systematic uncertainties that is 
required to achieve the optimal fit, which cannot be seen from these 
figures. It is noticeable that with the correction there is less shift in data 
relative to theory related to some of the correlated systematics that 
affect mainly the low $x$ and $Q^2$ data, e.g. procedural uncertainty 
$\delta_1$. Finally we show in Fig.~\ref{fig:PDFsfl} the effect this 
correction has on the PDFs obtained from the fit when it is included.  
These changes are seen to be very small, in particular for the global fit.
The change in the light sea for the HERA data only fit is due simply to
a reshuffling of quarks between different flavours, which is not constrained
in this type of fit. In practice the strange quark fraction increases.

In addition to a correction to $F_L$, we may also consider the effect on 
$F_2$. To do this we consider, as in~\cite{Martin:2003sk,ThorneFFNS}, 
a further correction
\begin{equation}
F_2(x,Q^2) \to F_2(x,Q^2)\left(1+\frac{a_i}{Q^2}\right)\;,
\end{equation}
where the $a_i$ correspond to $i=1,6$ bins in $x$, all below $x=0.01$, 
and are left free in the fit. This results in a small additional reduction of 
$\Delta\chi^2=10$ in the global fit, but with almost no effect at all on 
the comparison to the HERA data. Similarly it makes little difference in the 
HERA data only fit. It therefore appears that at the current level of 
accuracy the fit does not require any further corrections to $F_2$. Another 
possibility we consider is an additional $\propto 1/Q^4$ correction to $F_L$: 
this gives a very small further reduction of $\Delta\chi^2=5$, with no 
significant influence on the behaviour with $Q^2_{\rm min}$.

While it may be tempting to interpret the above result solely in terms of 
evidence for higher--twist corrections, it is important to emphasise that 
the contribution from $F_L$ is only significant at high $y=Q^2/sx$, and 
thus such a lower $Q^2$ correction is strongly correlated with low $x$. 
Indeed, if we instead try the correction
\begin{equation}\label{eq:flcorr1}
F_L^{(1)}(x,Q^2)=F_L(x,Q^2)\left(1+\frac{\alpha_S(Q^2)}{4\pi}\frac{b_1}{x^{b_2}}\right)\;,
\end{equation}
we find an reduction in $\Delta \chi^2=28$ with $b_1=0.014$ and $b_2=0.82$. 
However, as at fixed $y$ we have $ x \propto Q^2$, 
the power of $b_2 \lesssim 1$ in combination with the slow falling of 
$\alpha_S$ with $Q^2$ leads to the correction (\ref{eq:flcorr1}) being 
effectively $\sim 1/Q^2$ for fixed $y$, i.e. consistent with (\ref{eq:flcorr}).

Finally, we note that detailed examination of data against theory show that 
the theory predictions at high $Q^2$ and high $y$ show a tendency to 
undershoot the data, that is the opposite trend to the low $Q^2$ case; this means that for positive $b_1$ a smaller value of $b_2$ in 
(\ref{eq:flcorr1}) causes problems as it gives a negative correction to the cross section over a wide range of $x$ values, whereas  the high value of $b_2$ means the effect of the corrections is very much concentrated at small
$x$, i.e. only being significant for HERA data for small $Q^2$.
Indeed, if we try a $Q^2$ independent correction
\begin{equation}\label{eq:flcorr2}
F_L^{(1)}(x,Q^2)=F_L(x,Q^2)\left(1+c_1x^{c_2}\right)\;,
\end{equation}
then the best fit in fact results in an improvement of $\Delta \chi^2=13$, 
with $c_1=-1.97$ and $c_2=0.42$. This behaviour leads to a smaller predicted 
$F_L$, but has its main effect on high $y$ data at higher $x$ and 
therefore higher $Q^2$, reducing the tendency of the the theory to undershoot 
the data for the reduced cross section. Taking the sum of (\ref{eq:flcorr}) 
and  (\ref{eq:flcorr2}) allows an improvement in both the lower and higher 
$Q^2$ regions, and gives a reduction of $\Delta\chi^2=42$,  with 
$a=5.3\,{\rm GeV}^2$ and $c_1=-0.71$, $c_2=0.19$, with $a$ being somewhat 
higher than in the fit with only the $1/Q^2$ correction, consistent with 
there being some influence from the second term on the lower $x,Q^2$ region.

Hence, the ideal overall correction for $F_L$ is an increase at low $x$ 
and $Q^2$, of higher twist type, 
consistent  with the tendency for PDF predictions to undershoot 
the $F_L$ extraction from  \cite{Andreev:2013vha} for $Q^2 < 10~\GeV^2$, but 
a reduction at higher $x$ and $Q^2$. There are various possible mechanisms
where the value of $F_L$ obtained can be modified: the basic power-like 
higher twist type of correction explicitly considered; the effects of
absorptive corrections to evolution at small $x$ and $Q^2$; more general 
saturation corrections; and resummations of $\alpha_S\ln(1/x)$ terms in 
the perturbative series. A full study of these is beyond the scope of the 
present article.  Here we simply produce a parametric means of solving the 
most clear problem in the fit quality for the HERA data.

\section{Conclusions}\label{sec:conc}

We have examined the impact of the final HERA combination of inclusive cross 
section data presented in \cite{Abramowicz:2015mha}. We notice that we 
already predict these data very well with MMHT 2014 PDFs, particularly at NNLO,
and consequently their inclusion leads to very little impact on the central 
value of the MMHT2014 PDFs. The data do reduce the uncertainty in the PDFs,
mainly the gluon, though this is more noticeable in the uncertainty for 
predictions of benchmark LHC cross sections than in PDF plots, with the 
uncertainty on Higgs production via gluon fusion being reduced to about $90\%$
of the previous uncertainty. 
PDFs obtained from a fit to only the HERA combined data can vary significantly 
from those from the global fit for some PDFs, but most, including the gluon 
and down distributions, are similar to the global fit. There is 
very little constraint on antiquark flavour decomposition. The combined HERA 
data do seem to prefer a larger up quark above $x=0.2$, and this results in
a fit quality for $e^-$ charged current data in a HERA data only fit 
which is not reproducible in the global fit (though NNLO is better than NLO).  
We also confirm the result in \cite{Abramowicz:2015mha} that the 
fit quality improves with increasing $Q^2_{\min}$ (though our effect is 
smaller), and show that most of this effect is obtained just by changing 
the cut on the HERA data in the comparison, with little extra contribution 
when refitting is performed with the raised cut. We note that this $Q^2_{\min}$ behaviour can 
cured by the addition of a positive ``higher-twist'' like 
correction to $F_L$ and that this is more  effective than modifications to 
$F_2$. Small further improvements can also be achieved at higher $Q^2$ by
negative corrections to $F_L$ in this region. These corrections result in  
extremely little change in PDFs obtained from the fit.

Overall we conclude that the current PDFs, with very minor modifications, 
work extremely well for the final HERA data. The central values of the PDFs
are changed very little by the data, even if corrections are added to the 
theory to improve the fit quality. The data have an impact on 
uncertainties of PDFs obtained in the global fit, but very largely due to an 
improvement in the gluon uncertainty. LHC cross sections sensitive to this 
can  have a reduction in uncertainty to about $90~\%$ of their previous 
values. We do not deem this 
to be a significant enough effect to warrant an immediate new update of 
PDFs -- 
there is an ``uncertainty on the uncertainty'' which is very likely of this 
order. Instead we prefer to wait for a more substantial update which
will include the effects of e.g. full NNLO jet cross sections, NNLO 
corrections to differential top distributions~\cite{Czakon:2015owf}, 
and the inclusion of significantly more precise, varied, and higher energy
LHC data sets.

\section*{Acknowledgements}

We particularly thank W. J. Stirling  and G. Watt for numerous discussions on PDFs and for previous work without which this study would not be possible. This work is supported partly by the London Centre for Terauniverse Studies (LCTS), using funding from the European Research Council via the Advanced Investigator Grant 267352. RST would also like to thank the IPPP, Durham, for the award of a Research Associateship held while most of this work was performed. We thank the Science and Technology Facilities Council (STFC) for support via grant awards ST/J000515/1 and ST/L000377/1.

\bibliography{references}{}

\begin{thebibliography}{10}

\bibitem{MSTW}
A.~D. Martin, W.~J. Stirling, R.~S. Thorne, and G.~Watt,
\newblock Eur.Phys.J. {\bf C63}, 189 (2009), 0901.0002.

\bibitem{Harland-Lang:2014zoa}
L.~A. Harland-Lang, A.~D. Martin, P.~Motylinski, and R.~S. Thorne,
\newblock Eur. Phys. J. {\bf C75}, 204 (2015), 1412.3989.

\bibitem{ABM14}
S.~Alekhin, J.~Bluemlein, and S.~Moch,
\newblock Phys.Rev. {\bf D89}, 054028 (2014), 1310.3059.

\bibitem{JR14}
P.~Jimenez-Delgado and E.~Reya,
\newblock Phys.Rev. {\bf D89}, 074049 (2014), 1403.1852.

\bibitem{NNPDF3}
NNPDF, R.~D. Ball {\em et~al.},
\newblock JHEP {\bf 04}, 040 (2015), 1410.8849.

\bibitem{CT14}
S.~Dulat {\em et~al.},
\newblock Phys. Rev. {\bf D93}, 033006 (2016), 1506.07443.

\bibitem{Butterworth:2015oua}
J.~Butterworth {\em et~al.},
\newblock (2015), 1510.03865.

\bibitem{MMSTWW}
A.~D. Martin {\em et~al.},
\newblock Eur.Phys.J. {\bf C73}, 2318 (2013), 1211.1215.

\bibitem{Thorne}
R.~S. Thorne,
\newblock Phys.Rev. {\bf D86}, 074017 (2012), 1201.6180.

\bibitem{Bolton}
T.~Bolton,
\newblock (1997), hep-ex/9708014.

\bibitem{GGGP1}
A.~Gehrmann-De~Ridder, T.~Gehrmann, E.~Glover, and J.~Pires,
\newblock Phys.Rev.Lett. {\bf 110}, 162003 (2013), 1301.7310.

\bibitem{GGGP2}
J.~Currie, A.~Gehrmann-De~Ridder, E.~Glover, and J.~Pires,
\newblock JHEP {\bf 1401}, 110 (2014), 1310.3993.

\bibitem{H1+ZEUS}
H1 and ZEUS Collaboration, F.~Aaron {\em et~al.},
\newblock JHEP {\bf 1001}, 109 (2010), 0911.0884.

\bibitem{H1+ZEUScharm}
H1 Collaboration, ZEUS Collaboration, H.~Abramowicz {\em et~al.},
\newblock Eur.Phys.J. {\bf C73}, 2311 (2013), 1211.1182.

\bibitem{Abramowicz:2015mha}
ZEUS, H1, H.~Abramowicz {\em et~al.},
\newblock (2015), 1506.06042.

\bibitem{Thorne:2015caa}
R.~S. Thorne, L.~A. Harland-Lang, A.~D. Martin, and P.~Motylinski,
\newblock {The Effect of Final HERA inclusive Cross Section Data on MMHT2014
  PDFs},
\newblock in {\em {Proceedings, 2015 European Physical Society Conference on
  High Energy Physics (EPS-HEP 2015)}}, 2015, 1508.06621.

\bibitem{Rojo:2015nxa}
J.~Rojo,
\newblock {Progress in the NNPDF global analysis and the impact of the legacy
  HERA combination},
\newblock in {\em {Proceedings, 2015 European Physical Society Conference on
  High Energy Physics (EPS-HEP 2015)}}, 2015, 1508.07731.

\bibitem{MSTWDIS}
R.~Thorne, A.~Martin, W.~Stirling, and G.~Watt,
\newblock PoS {\bf DIS2010}, 052 (2010), 1006.2753.

\bibitem{WZNNLO}
R.~Hamberg, W.~van Neerven, and T.~Matsuura,
\newblock Nucl.Phys. {\bf B359}, 343 (1991).

\bibitem{HiggsNNLO1}
R.~V. Harlander and W.~B. Kilgore,
\newblock Phys.Rev.Lett. {\bf 88}, 201801 (2002), hep-ph/0201206.

\bibitem{HiggsNNLO2}
A.~Djouadi, M.~Spira, and P.~Zerwas,
\newblock Phys.Lett. {\bf B264}, 440 (1991).

\bibitem{topNNLO}
M.~Czakon, P.~Fiedler, and A.~Mitov,
\newblock Phys.Rev.Lett. {\bf 110}, 252004 (2013), 1303.6254.

\bibitem{Martin:2003sk}
A.~D. Martin, R.~G. Roberts, W.~J. Stirling, and R.~S. Thorne,
\newblock Eur. Phys. J. {\bf C35}, 325 (2004), hep-ph/0308087.

\bibitem{ThorneFFNS}
R.~Thorne,
\newblock Eur.Phys.J. {\bf C74}, 2958 (2014), 1402.3536.

\bibitem{Andreev:2013vha}
H1, V.~Andreev {\em et~al.},
\newblock Eur. Phys. J. {\bf C74}, 2814 (2014), 1312.4821.

\bibitem{Czakon:2015owf}
M.~Czakon, D.~Heymes, and A.~Mitov,
\newblock Phys. Rev. Lett. {\bf 116}, 082003 (2016), 1511.00549.

\end{thebibliography}
\bibliographystyle{h-physrev}

\end{document}